\documentclass[prb,twocolumn,amsmath,amssymb,floatfix,superscriptaddress,longbibliography,nofootinbib,reprint]{revtex4-2}

\usepackage{graphicx} 
\usepackage{bm} 

\usepackage{hyperref}

\usepackage{braket,bbold,color,enumerate}

\usepackage{geometry}
\usepackage{amsmath,amssymb,xcolor}
\usepackage{amsfonts}
\usepackage{amsmath}
\usepackage{mathtools}
\usepackage{dsfont}

\newcommand{\obs}{\mathcal{O}}

\newcommand{\im}{\mathrm{i}}
\newcommand{\id}{\mathbb{1}}
\newcommand{\CZ}{\mathrm{CZ}}

\newcommand{\CZX}{\mathrm{CZX}}

\newcommand{\HPM}{H_\mathrm{PM}}
\newcommand{\UCZX}{U_\mathrm{CZX}}

\newcommand{\PM}{\mathrm{PM}}

\usepackage{xcolor}

\begin{document}

\title{Agnostic and efficient detection of $\mathbb{Z}_2$ symmetry-protected topological phases \\
in projected entangled-pair states}

\author{Alexandre Chaduteau}
\thanks{These authors contributed equally to this work.}
\affiliation{Blackett Laboratory, Imperial College London, London SW7 2AZ, United Kingdom}

\author{Hao Chen}
\thanks{These authors contributed equally to this work.}
\affiliation{Department of Physics, University of Zürich, 8057 Zürich, Switzerland}

\author{Juraj Hasik}
\affiliation{Department of Physics, University of Zürich, 8057 Zürich, Switzerland}

\author{Titus Neupert}
\affiliation{Department of Physics, University of Zürich, 8057 Zürich, Switzerland}

\begin{abstract}
Symmetry-protected topological (SPT) phases are the simplest example of quantum phases that cannot be detected by local order parameters. Their classification is by now well-established from multiple perspectives, a prominent one being via tensor network states. The practical question, however, is not mere classification, but detection: given the ground state of an interacting Hamiltonian and a symmetry group $G$ along with its representation, 
is there an efficient way to tell if the state is in a SPT phase with respect to $G$? While there exist such practical tests in one spatial dimension (1D), methods in 2D have so far been restricted to finite-size systems or require an explicit form of virtual symmetries of the tensor networks representing the ground state. Here, we develop an agnostic and efficient method to detect interacting $\mathbb{Z}_2$ 2D SPT phases, using an infinite projected entangled-pair (PEPS) representation of the ground state. The method is agnostic to the symmetry representation in the virtual space of the PEPS tensor and applicable independent of the bond dimension of the tensors. Moreover, it is efficient as it does not require finite-size scalings: it works directly in the thermodynamic limit. We numerically demonstrate the procedure by mapping out phase boundaries away from fixed-point wavefunctions. We extract the SPT label cocycle data explicitly from the PEPS tensor.
\end{abstract}

\maketitle


\section{Introduction}
Conventional quantum phases, represented by ground states of Hamiltonians, are distinguished by a local order parameter such as magnetization in the paradigmatic Ising model. In recent decades, however, topological phases~\cite{RevModPhys.89.041004} have been found to evade this simple diagnostic, in that they lack a local order parameter and are not distinguished by broken symmetries. These phases are instead distinguished by topological invariants, which are robust against local perturbations, quantized, and only change at a topological phase transition. The simplest such topological phases are symmetry-protected topological (SPT) phases~\cite{Senthil_2015}, which are adiabatically connected to a trivial product state if the symmetry requirement is not enforced. Well-known non-interacting examples are topological insulators~\cite{RevModPhys.82.3045}, for which full classification tables form a well-established part of the literature. As a natural extension, interacting SPTs have been thoroughly studied~\cite{doi:10.1126/science.1227224, PhysRevB.90.245120, sohal2026symmetrytwistedmultientropiesorderparameters, PhysRevX.10.031055} and experimentally observed~\cite{Experimental_interacting_SPT_2019}. In one dimension~(1D), interacting SPTs have been exhaustively classified through the tools of matrix product states~(MPSs)~\cite{PhysRevB.83.075103, Chen_Gu_Wen_SPT_2013, schuch_classif_phases_MPS_PEPS, PhysRevB.96.165124}. The culminating statement is that the possible bosonic SPT phases for a symmetry group $G$ are classified by the second cohomology group $\mathcal{H}^2(G,U(1))$. A similar treatment using projected entangled-pair states (PEPS) provided a (partial) classification of bosonic SPTs in general $d$ dimensions~\cite{CZX_model_paper, Chen_Gu_Wen_SPT_2013} in terms of higher-order cohomology groups $\mathcal{H}^{d+1}(G,U(1))$. This was later generalized by more sophisticated tools such as cobordism theory~\cite{kapustin2014symmetryprotectedtopologicalphases, PhysRevX.3.011016}.
The focus on PEPS is natural for the 2D problem, as these constitute a systematically improvable space for short-range entangled states and underlie one of the leading numerical methods for computing ground states of interacting 2D quantum systems. 

The more physical question remains: given the ground state $\ket{\psi}$ of a local Hamiltonian which is symmetric under a group $G$, is there a practical method to tell whether $\ket{\psi}$ is in a SPT phase? Specifically, this method should work for interacting Hamiltonians, for which there are no readily available single-particle quantities and topological invariants computed from them~\cite{doi:10.1126/science.1227224, PhysRevB.76.045302}. 
\begin{figure*}
    \centering
    \includegraphics[width=0.9\linewidth]{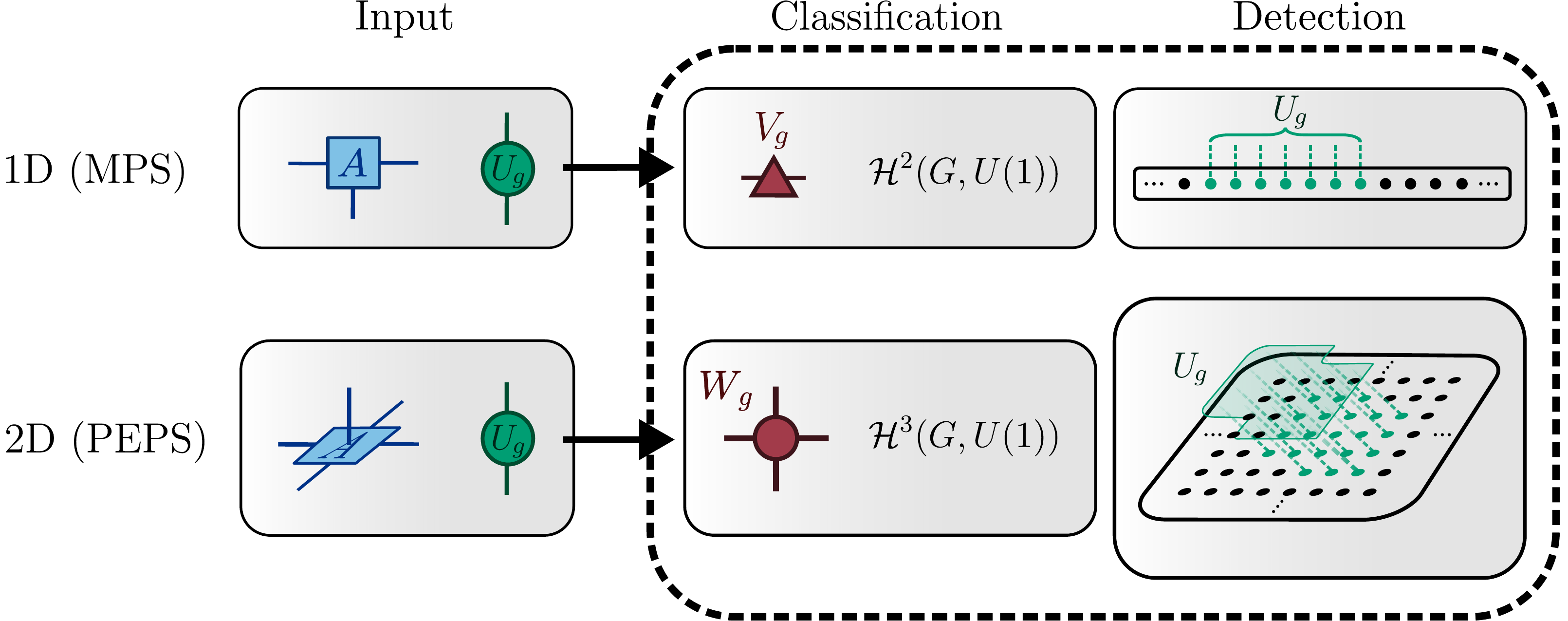}
    \caption{Overview of tensor network methods for SPT classification and agnostic detection in 1D and 2D. In 1D, the input is a (translationally invariant) representation of the ground state $\ket{\psi}$ with an (injective) MPS tensor, as well as the protecting symmetry group $G$ and its unitary representation on the physical leg $U_g$. The realized SPT can be identified by directly extracting the virtual symmetry action $V_g$ and computing its 2-cocycle [see Sec.~\ref{sec:1D_section} and Ref.~\onlinecite{PhysRevB.86.125441}]. Alternatively, the SPT phase can also be identified without $V_g$ via string-order parameters, which compute the expectation value of applying a symmetry operation $U_g$ to a segment (potentially with charged operators at the two edges). Similarly, in 2D (see Sec.~\ref{sec:2D_section}), the input is a (MPO-injective) PEPS tensor and the symmetry operations $U_g$. The classification is obtained from the corresponding virtual action $W_g$. Our main result is a numerically efficient method to compute the expectation value of the symmetry action $U_g$ on a subregion, enabling us to probe 2D $\mathbb{Z}_2$ SPT phases while avoiding the computation of the virtual symmetry action $W_g$.}
    \label{fig:summary_figure}
\end{figure*}
The answer lies in the projective decomposition of the representation of the symmetry operations. In 1D, a non-trivial SPT phase exhibits a projective representation of the symmetry group at the edge~\cite{PhysRevB.90.235137, PhysRevB.96.165124}, described by an element $[\omega] \in \mathcal{H}^2(G,U(1))$, called a $2$-cocycle. Similarly, in 2D, a non-trivial SPT phase has a non-trivial $3$-cocycle $[\omega] \in \mathcal{H}^3(G,U(1))$, which is related to the associativity of the low-energy effective symmetry actions at the 1D boundaries~\cite{CZX_model_paper, Chen_Gu_Wen_SPT_2013}. In the tensor network representation of SPT states, this effective symmetry can be found as \emph{virtual} symmetry, acting only on the virtual Hilbert space of the tensor network~\cite{RevModPhys.93.045003}. One method to detect SPTs therefore involves finding the explicit virtual symmetry representation and directly computing cocycles. In 1D this can be done using MPS~\cite{PhysRevB.86.125441}, where virtual symmetries have  matrix representations. Instead, in the 2D case such virtual symmetries, now represented by matrix product operators (MPOs) acting on virtual degrees of freedom of PEPS states, are not known to be easily computable; previous works~\cite{Sahinoglu2021, PhysRevB.94.205150} have implicitly imposed these virtual symmetries on the PEPS tensors. With practical applications in mind, this approach remains limited, since the explicit form of the virtual symmetry MPOs is not readily accessible in generic PEPS optimizations.

We are then led to the question: can we detect 2D SPT phases in an agnostic way, i.e. without relying on explicit knowledge of virtual symmetries? In other words, are there physical observables that are sensitive to the non-trivial cocycles described above? The answer is yes: the key idea is to insert ``symmetry defects" in the system and to construct observables that are sensitive to such defects. In 1D, these are string-order parameters and their variants~\cite{PhysRevB.86.125441, PhysRevLett.100.167202, PhysRevLett.109.050402}. Our main result, summarized in Fig.~\ref{fig:summary_figure}, is a computationally efficient extension of this approach to PEPS-represented 2D $\mathbb{Z}_2$ SPT states.

We base our analysis on expectation values of symmetries applied to a subregion of the system, an established route to extract information about projective symmetry algebras~\cite{PhysRevB.90.235113, PhysRevB.90.235137, PhysRevLett.118.216402, mfwg-svcp, PhysRevB.3.3918, Fradkin2017, PhysRevB.91.195134, PhysRevB.106.094415}.
The starting point for our method is infinite-PEPS (iPEPS) representation of the state. Previously proposed diagnostics~\cite{PhysRevB.86.125441, PhysRevB.90.235113, PhysRevLett.118.216402, PhysRevB.93.155163, mfwg-svcp, PhysRevB.94.205150, PhysRevLett.120.156601, You2014, PhysRevB.95.045111, PhysRevLett.118.216402, PhysRevB.90.235137, Xu_2026, PhysRevB.106.094415, PhysRevB.90.205114} were performed on finite-size systems, generally considering a cylinder of finite width; a scaling analysis against the size was required. By contrast, our diagnostic works directly in the thermodynamic limit, taking advantage of the iPEPS structure.
 There, for gapped non-chiral states, the exact expectation values are dictated only by leading eigenvectors of transfer operators. This reduces $d$-dimensional overlaps to $(d-1)$-dimensional contractions, without the limitations of the cylinder geometry.

Our implementation of the diagnostic employs the corner transfer matrix renormalization group (CTMRG) algorithm to efficiently compute leading eigenvectors of iPEPS transfer operators. Combining this with iPEPS optimizations, we are able to map out the entire interpolation between a trivial and non-trivial SPT phase.

The organization of the paper is as follows. In Sec.~\ref{sec:1D_section} we review (bosonic) 1D SPTs and their methods of detection, both those that rely on the knowledge of virtual symmetries and those that do not. In Sec.~\ref{sec:2D_section}, we treat the 2D case and develop a general method to compute observables that are sensitive to non-trivial (bosonic) SPT phases. We apply this technique in Sec.~\ref{sec:example} to the simplest 2D SPT protected by an on-site $\mathbb{Z}_2$ symmetry group~\cite{CZX_model_paper, PhysRevB.94.205150, PhysRevB.91.195134, PhysRevB.96.195101}. We finally conclude in Sec.~\ref{sec:conclusion} and discuss potential extensions.

\section{One-dimensional SPTs: review}\label{sec:1D_section}
We begin with a concise review of MPS methods for detecting 1D SPTs~\cite{PhysRevB.86.125441}.
\begin{figure}
    \centering
    \includegraphics[width=\linewidth]{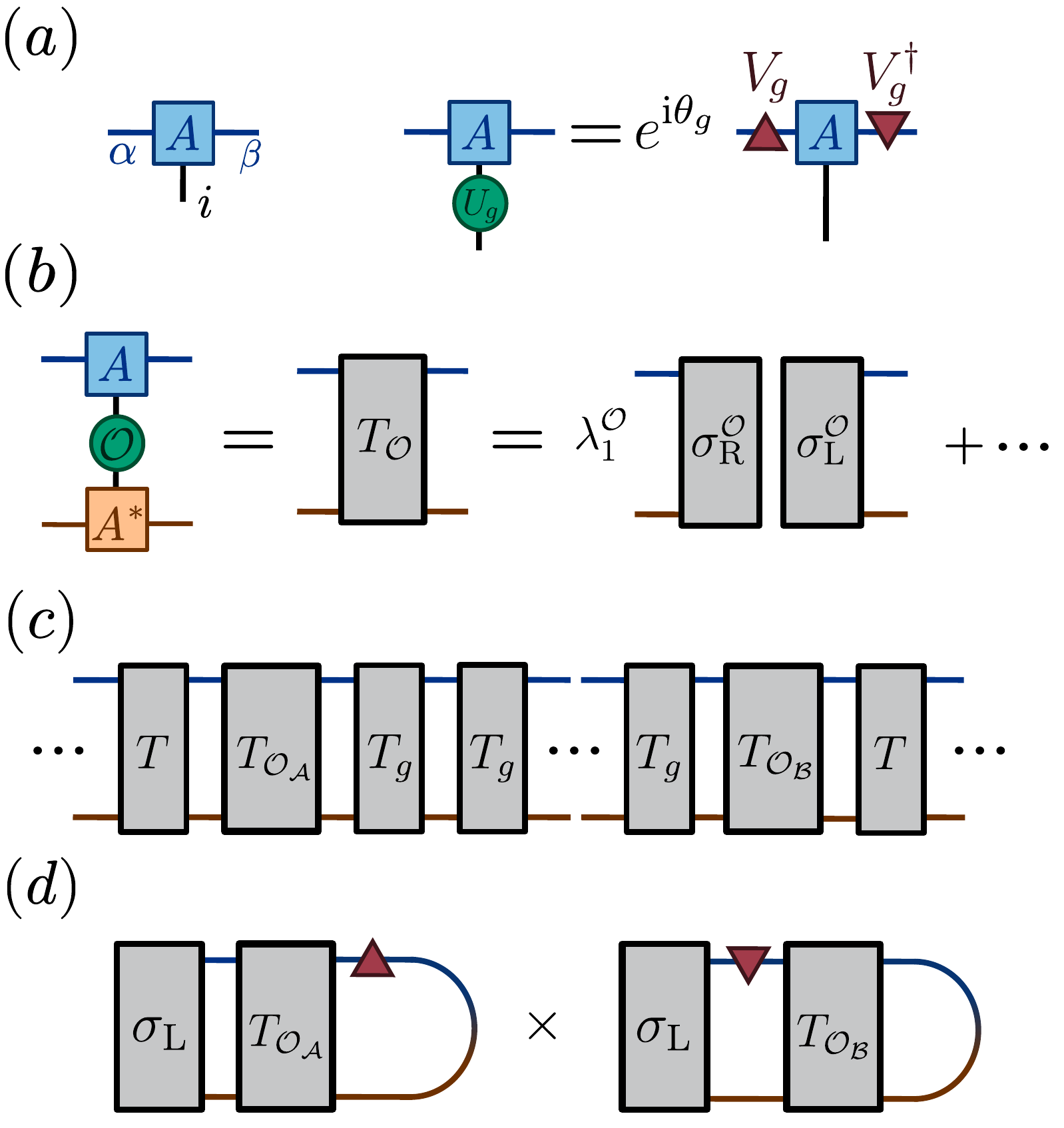}
    \caption{Diagrammatic representations of contractions for infinite MPSs~\cite{PhysRevB.86.125441, RevModPhys.93.045003}. (a) We represent the MPS tensor $A^i_{\alpha \beta}$ and its indices with horizontal virtual legs $\alpha,\beta$ and vertical physical legs $i$. With this notation, Eq.~\eqref{eq:pulling_through_condition_1D} is as shown. (b) Transfer matrix $T_\obs$ of Eq.~\eqref{eq:transfer_matrix_definition_1D} and its eigenvector decomposition, with only the projector onto the first eigenvalue shown. (c) The string-order parameter $\mathcal{S}(g,\obs_A,\obs_B)$ of Eq.~\eqref{eq:string_order_parameter}, as a contraction of transfer matrices. (d) Simplification of the string-order parameter as a $0$D contraction, using the (unique) left/right leading eigenvectors $\sigma_{\mathrm{L},\mathrm{R}}$ of $T$ and the cancellation $V_gV^\dagger_g = \id$ in the bulk. We also used $\sigma_\mathrm{R}=\id$ for right-canonical MPS~\cite{PhysRevLett.100.167202}.
    We also ignored the overall phase factor due to the presence of multiple $e^{\im \theta_g}$ from (a).}
    \label{fig:transfer_matrices_1D}
\end{figure}
Consider a 1D spin chain of infinite length, with basis $\{\ket{\ldots ,i_n, i_{n+1}, \ldots}\}$, with each $i$ going from $1$ to the local Hilbert space dimension $p$.
The ground state of a gapped, local Hamiltonian for this system can be arbitrarily well-approximated by injective MPS~\cite{PhysRevB.73.094423, 8589272,RevModPhys.93.045003}, which we will define shortly. 
Furthermore, assuming translational invariance, such a ground state can be fully described by a MPS generated by a single tensor $A^i_{\alpha,\beta}$ as
\begin{equation}
\label{eq:1}
    \ket{\psi} = \sum_{\{i\}}\left[\ldots A^{i_{n}}A^{i_{n+1}}\ldots\right] \ket{\ldots i_{n}, i_{n+1} \ldots},
\end{equation}
where the sum goes over all possible spin configurations $\{i\}$. For injective MPS, boundary conditions at infinity do not affect bulk expectation values. The virtual indices $\alpha$ and $\beta$ of the MPS tensor range from $1$ to a specified bond dimension $\chi$.
Next, we define generalized MPS transfer matrices (TMs)  
\begin{equation}\label{eq:transfer_matrix_definition_1D}
    T_{\obs} := \sum_{ij} \obs_{ij}A^j \otimes (A^i)^*,
\end{equation}
for $\obs_{ij}$ the matrix elements of some on-site physical operator [see Fig.~\ref{fig:transfer_matrices_1D}(b)]. Importantly, for $\obs=\id$, we have the ``regular" TM $T = \sum_i A^i \otimes (A^i)^*$. The MPS is injective if and only if $T$ has a unique leading eigenvalue $\lambda_1$, with corresponding right and left eigenvector pair $\sigma_\mathrm{R}$, $\sigma_\mathrm{L}$. Rescaling $A\rightarrow A/\sqrt{\lambda_1}$ and normalizing the overlap of $\sigma_\mathrm{L}$ and $\sigma_\mathrm{R}$ to one, we have $\langle\psi|\psi\rangle = 1$, and all expectation values are determined by these leading eigenvectors.

We consider a finite group $G$ to be a global symmetry of the system, with each $g\in G$ unitarily represented as
$U^{\otimes}_g = \bigotimes_{n} U_g$, where $U_g$ is the on-site action. For a MPS [Eq.~\eqref{eq:1}] symmetric under $G$, the fundamental theorem of MPS~\cite{PhysRevLett.100.167202, CIRAC2017100} establishes that the tensor $A$ must transform under the action $U_g$ by  
\begin{equation}
\label{eq:pulling_through_condition_1D}
  \sum_{j}(U_g)_{ij}A^j = e^{\im \theta_g}V_g A^i V_g^\dagger,
\end{equation}
as shown schematically in Fig.~\ref{fig:transfer_matrices_1D}(a). We call $V_g$ the virtual representation of the symmetry action, which can be chosen unitary for injective MPS. The phase factor $\theta_g$ comes from the $U(1)$ phase ambiguity in general wavefunctions. Considering the generalized TM $T_{U_g}\equiv T_g$ with $\obs = U_g$, the virtual representations $V_g$ can be obtained from leading eigenvectors of $T_g$~\cite{PhysRevB.86.125441}. More precisely, for a right-canonical MPS, where $\sigma_\mathrm{R} = \id$, the leading right eigenvector $\sigma^g_\mathrm{R}$ of $T_g$ is just $V_g$. Thus, diagonalizing $|G|$ mixed TMs yields the full virtual representation $\{V_g\}$ of $G$, from which one can explicitly determine the 2-cocycle $\left[ \omega \right] \in \mathcal{H}^2(G,U(1))$ that labels the SPT phase~\cite{Chen_Gu_Wen_SPT_2013}.

On the other hand, as per our initial motivation (and with the imminent 2D case in mind), can we detect the SPT phase \emph{without} knowing $V_g$? The string-order parameter (SOP)~\cite{PhysRevB.86.125441, PhysRevLett.100.167202} is an affirmative answer to this question. It inserts an extended symmetry defect by applying the symmetry operation on all sites $n=2,\ldots, l-1$, where $l$ is the length of the defect. Its sensitivity to the projective representation of the symmetry depends on the choice of operators $\obs_A, \obs_B$ placed at the ends of the symmetry defect. Explicitly, the SOP $\mathcal{S}(g, \obs_A, \obs_B)$ is defined as
\begin{equation}\label{eq:string_order_parameter}
\begin{aligned}
    \lim_{l\to\infty}\bra{\psi}\obs_A(1)\left(\prod_{n=2}^{l-1}U_g\right)\obs_B(l)\ket{\psi},
\end{aligned}
\end{equation}
which can be evaluated via TMs as in Fig.~\ref{fig:transfer_matrices_1D}(c). By the above definitions, $\mathcal{S}(g, \obs_A, \obs_B)$ reduces to a product of $0$D contractions involving the virtual symmetry action $V_g$, as shown in Fig.~\ref{fig:transfer_matrices_1D}(d). In this infinite chain, the operators $\obs_A$ and $\obs_B$ play the role of boundary conditions.

For phases that are not symmetric under $U_g$, the SOP is generically $0$. Within symmetric phases, however, the SOP can distinguish certain trivial and SPT phases. Its value depends on symmetry quantum numbers carried by the endpoint operators $\obs_A$ and $\obs_B$. For example, consider commuting symmetries $g,h$, for which the virtual representations satisfy
\begin{equation}
V_gV_hV_g^\dagger V_h^\dagger
=
\beta_{gh}
=
\frac{\omega(g,h)}{\omega(h,g)},
\end{equation}
where $\omega$ is the $2$-cocycle characterizing the SPT phase. In Ref.~\onlinecite{PhysRevB.86.125441} it was shown that the string order parameter obeys a selection rule: if $\obs_A$ and $\obs_B$ transform under $h$ with character $\chi(h)$, then $\mathcal{S}(g,\obs_A,\obs_B)$ can be nonzero only when $\chi(h)=\beta_{gh}$. For instance, when $G=\mathbb{Z}_2\times\mathbb{Z}_2$, relevant to the spin-$1$ Heisenberg chain, the non-trivial SPT phase has $\beta_{gh}=-1$, while the trivial phase has $\beta_{gh}=1$. The corresponding selection rule therefore distinguishes the two phases and directly probes the invariant $\beta_{gh}$.

While additional subtleties arise for larger symmetry groups, as well as for spatial and anti-unitary symmetries~\cite{PhysRevB.86.125441, PhysRevLett.100.167202, pollmann_entanglement_1d_2010, PhysRevLett.109.050402, turner_pollmann_1d_top_fermions_entanglement}, the underlying idea remains the same: SOPs provide a partial characterization of SPT phases through selection rules relating the symmetry characters of endpoint operators to the projective symmetry representation on the virtual space.
Moreover, the reduction from an infinite 1D contraction to a 0D contraction enables this method to be efficient.

Another hallmark signature that has been used to detect SPT phases is the entanglement spectrum (ES)~\cite{original_entang_spectrum_haldane, pollmann_entanglement_1d_2010}. Although agnostic, the ES does not provide information as to which symmetry can protect a non-trivial SPT phase and which projective representation is realized in the phase.

\section{Two-dimensional SPTs}\label{sec:2D_section}
With the 1D case reviewed, we now move on to 2D SPTs. Our aim will be to study the action of the protecting symmetry $G$ only on part of the 2D system. First, we consider cylindrical geometries and treat the action via transfer matrices of the tensor network on cylinders and then consider the infinite-width cylinder limit, where a truly 2D geometry is recovered.

\subsection{PEPS, transfer matrices and $\tau_g$}
\begin{figure}
    \centering
    \includegraphics[width=\linewidth]{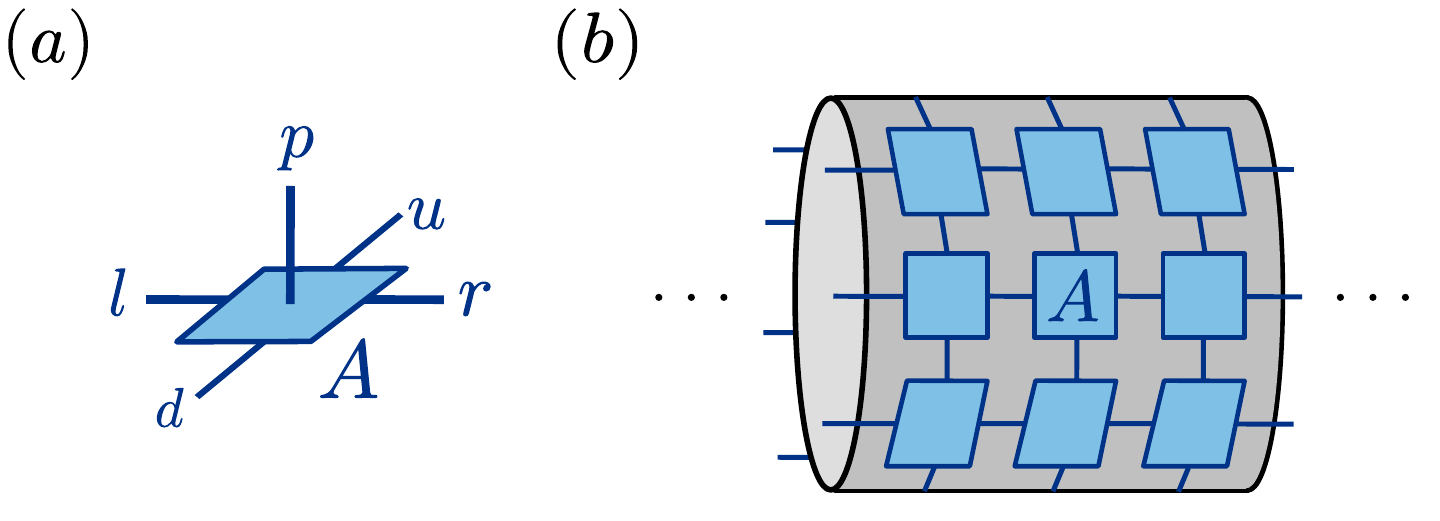}
    \caption{(a) Diagrammatic definition of the PEPS tensor used to describe 2D ground states. There are four virtual legs $u,r,d,l$ (of bond dimension $D$) and one physical leg $p$. (b) On a cylinder geometry, a PEPS wavefunction is obtained by tessellating the cylinder with tensors $A$, with neighbouring tensors contracted via their virtual legs as shown here for a width of $w=7$. Depending on the phase and the choice of boundary conditions, the resulting state can correspond to a particular ground state. For clarity, physical legs are omitted.}
    \label{fig:PEPS_cylinder_geometry}
\end{figure}

We first introduce protagonist quantities involved in the detection method. Following the introduction for 1D systems, we consider gapped local Hamiltonians defined on a 2D Euclidean lattice with a spin degree of freedom on each site $i$. We focus on translationally invariant Hamiltonians\footnote{The discussion can be adapted to the non-translationally-invariant case by blocking together the tensors of multiple sites.}, with ground state $|\psi\rangle$ described by a single PEPS tensor $A^p_{urdl}$, as shown in Fig.~\ref{fig:PEPS_cylinder_geometry}(a). The physical index is $p$ and ($urdl$) are four virtual indices corresponding to up, right, down, and left directions on a square lattice, each with bond dimension $D$. The ground state is obtained by tessellating the lattice with tensors $A$ and contracting them by connected virtual indices.

We again consider a Hamiltonian symmetric under a finite group $G$, represented by global unitary symmetries $U_g$. Let us consider the action of $g\in G$ restricted to a (2D) subregion $\Sigma$ with a (1D) boundary $\partial\Sigma$ of length $|\partial\Sigma|=L$, represented as $U^\Sigma_g = \otimes_{i\in \Sigma}U_g$. Importantly,  $U^\Sigma_g$ is not a symmetry of the system; based on previous works~\cite{PhysRevB.106.094415, PhysRevB.90.235113, PhysRevB.90.235137, PhysRevLett.118.216402, Xu_2026, mfwg-svcp, PhysRevB.111.205104}, we stipulate that the ground-state expectation value $\langle U^\Sigma_g \rangle$ obeys the scaling relation
\begin{equation}\label{eq:partial_symmetry_action_scaling}
\langle U^\Sigma_g \rangle \approx e^{-\alpha_g L}f(L)
\end{equation}
for $L \gg \xi$ where $\xi$ is the correlation length of the ground state. Here $\alpha_g$ is a non-universal constant; the function $f(L)$ is some  envelope (usually involving trigonometric functions) that does not decay with $L$. Its amplitude has been studied as a topological term~\cite{PhysRevB.106.094415, Xu_2026, mfwg-svcp}, present only in certain non-trivial topological phases protected by $G$. However, this method is not applicable to the simplest and thus most relevant SPTs [see Appendix~\ref{appendix:TQFT}]. 

\begin{figure}
    \centering
    \includegraphics[width=\linewidth]{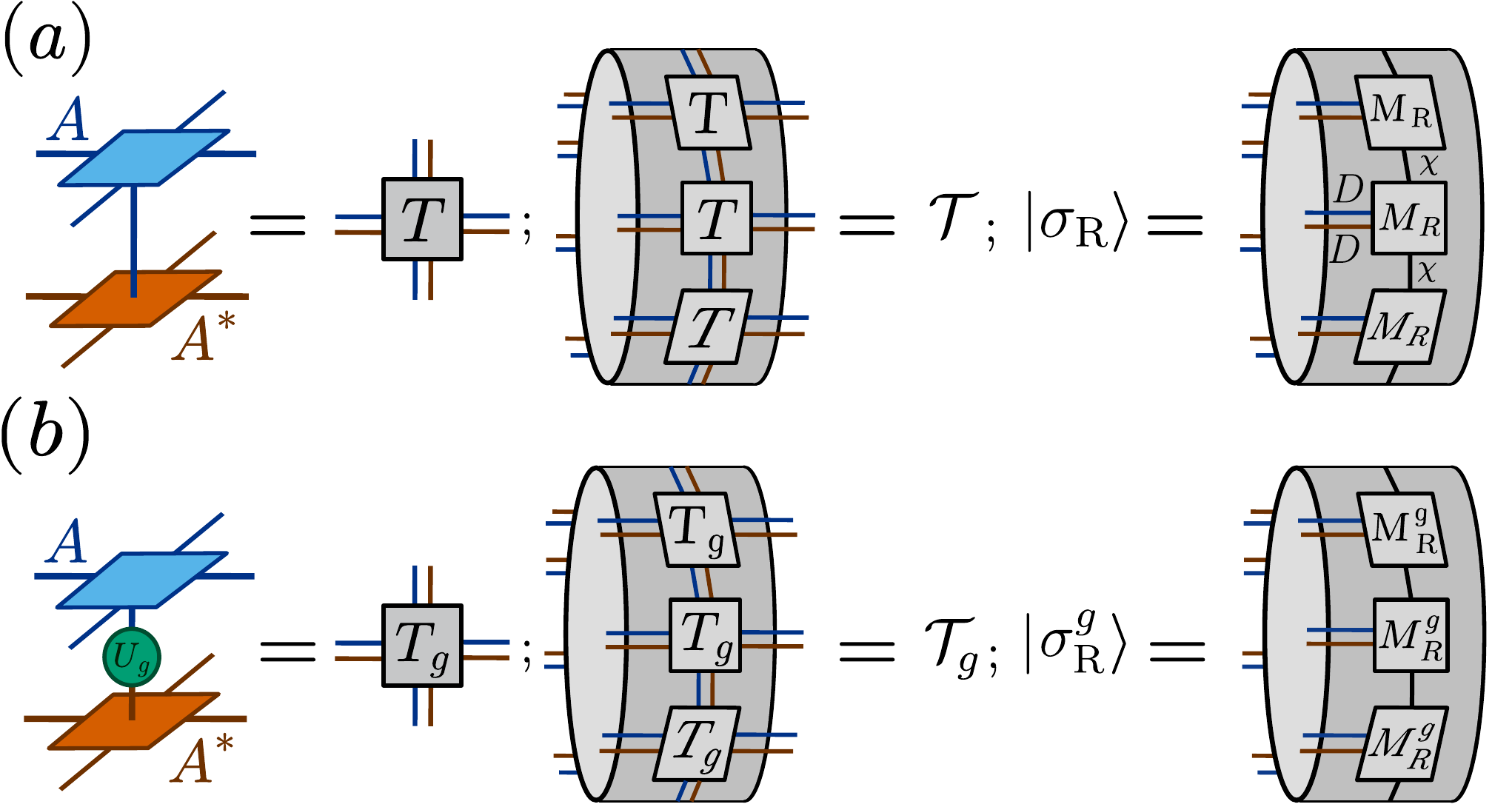}
    \caption{(a) A regular cylinder transfer matrix $\mathcal{T}$ constructed from the double-layer PEPS tensor $T$ by contracting a width-$L$ column of $T$ tensors. Its leading right-eigenvector $\ket{\sigma_\mathrm{R}}$, expressed as a uniform MPS generated by single tensor $M_\mathrm{R}$; (b) generalized cylinder transfer matrix $\mathcal{T}_{U_g}\equiv\mathcal{T}_g$ obtained by contracting a width-$L$ column of double-layer PEPS tensors $T_{U_g}\equiv T_g$ with operator $\mathcal{O}=U_g$ acting on the physical legs. Its leading right-eigenvector $\ket{\sigma^g_\mathrm{R}}$ as uniform MPS generated by tensor $M^g_\mathrm{R}$.}
    \label{fig:transfer_matrices_2D}
\end{figure}

\begin{figure}
    \centering
    \includegraphics[width=0.9\linewidth]{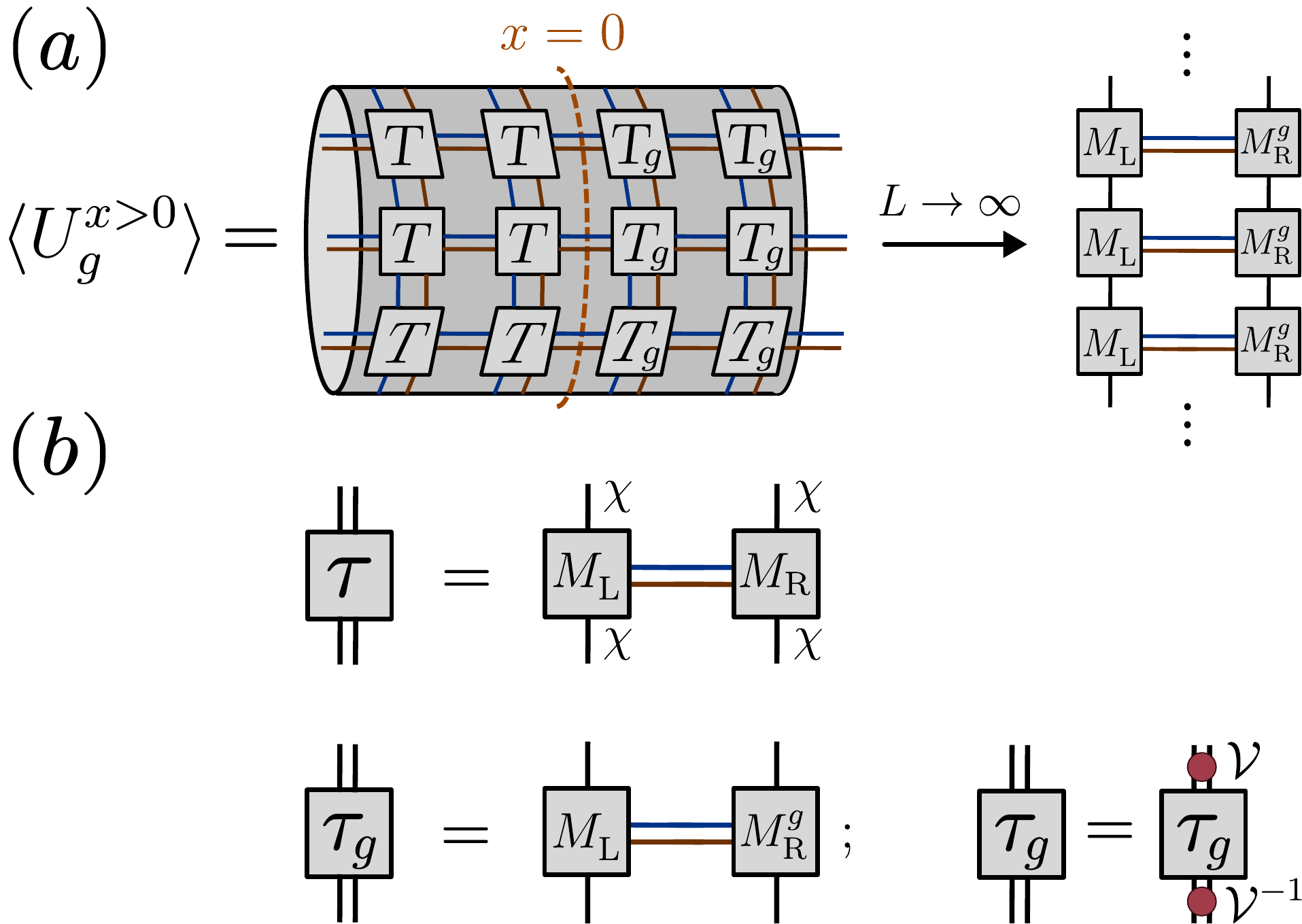}
    \caption{(a) The partial symmetry operation expectation value $\langle U^{x>0}_g\rangle$ as an infinite contraction of double-layer PEPS tensors $T$ for $x < 0$ and $T_g$ for $x>0$ (in the diagram, it is understood that the tensors repeat infinitely in the horizontal direction).
    By the definitions of Fig.~\ref{fig:transfer_matrices_2D}, this corresponds to the overlap $\bra{\sigma_\mathrm{L}}\sigma^g_\mathrm{R}\rangle$, which is further expressed as contraction of their respective boundary MPSs.
    (b) Definitions of the normalization single-site tensor $\tau$ dictating $\langle \id \rangle = \bra{\psi}\psi\rangle$, as well as the single-site tensor $\tau_g$ dictating $\langle U^{x>0}_g \rangle$ in (a). These are tensors with each leg of environment bond dimension $\chi$. Our main result is an anti-linear symmetry $\mathcal{V}$ of $\tau_g$ [see Eq.~\eqref{eq:tau_g_Z_2_case_constraint1}], from which we can extract the $\mathbb{Z}_2$ SPT label numerically via $\mathcal{V}^2=\omega(g,g,g)\id$.}
        \label{fig:partial_sym_operation_without_W}
\end{figure}

Let us set up a particularly simple way to probe the scaling Eq.~\eqref{eq:partial_symmetry_action_scaling}. We consider an infinite cylinder of width $L$ with open boundary conditions in the (horizontal) $x$ direction and periodic boundary conditions (PBC) in the (vertical) $y$ direction, shown in Fig.~\ref{fig:PEPS_cylinder_geometry}(b). We study the action of a symmetry $g$ restricted to the right half of this cylinder, i.e. $U^{x>0}_g$. Leveraging the PEPS structure, we can express $\langle U^{x>0}_g \rangle$ conveniently through generalized cylinder TMs $\mathcal{T}_\mathcal{O}$ defined in Fig.~\ref{fig:transfer_matrices_2D} as 
\begin{equation}
    \langle U^{x>0}_g \rangle = \prod_{x=-\infty}^0\mathcal{T}\prod^{\infty}_{x=1}\mathcal{T}_g,
\end{equation}
where $\mathcal{T}_\id=\mathcal{T}$ is the regular cylinder TM, which also gives the norm of the PEPS as $\langle\psi|\psi\rangle=\prod^{\infty}_{x=-\infty}\mathcal{T}$. 
For both trivial/nontrivial SPT phases, $\mathcal{T}$ has a unique leading eigenvalue, which can be normalized to 1.
The corresponding leading right and left eigenvectors are $\ket{\sigma_{\mathrm{R}}}$ and $\bra{\sigma_{\mathrm{L}}}$ respectively, which we assume do not spontaneously break translational symmetry along the $y$ direction. As they contract with right/left virtual legs of $\mathcal{T}$, they live in a ``virtual" Hilbert space of dimension $D^{2L}$.
The leading eigenvector $\ket{\sigma_\mathrm{R}}$ admits a representation as a translationally invariant MPS, whose single-site tensor is $(M_\mathrm{R})^{\alpha,\beta}_{a,b}$. 
The virtual indices $\alpha,\beta \in \{1,\ldots,D\}$ are inherited from the PEPS tensor's virtual indices; its internal indices $a,b\in \{1,\ldots,\chi\}$ run up to an environment dimension $\chi$ [see Fig.~\ref{fig:transfer_matrices_2D}(a)]. Later we will also use its matricized version $\sigma_{\mathrm{R}}$, which is a matrix product operator (MPO) with size $D^L \times D^L$~\cite{RevModPhys.93.045003}.
Analogously, $\bra{\sigma_\mathrm{L}}$ admits an MPS representation with single-site tensor $M_\mathrm{L}$. 
Furthermore, we assume $\mathcal{T}_g$ also has unique leading right and left eigenvectors $\ket{\sigma^g_\mathrm{R}}$ and $\bra{\sigma^g_\mathrm{L}}$, which can be expressed as single-site MPSs generated by tensors $M^g_{\mathrm{R}}$ and $M^g_{\mathrm{L}}$, respectively. The expectation value $\langle U^{x>0}_g\rangle$ on PEPS thus simplifies to 
\begin{equation}
    \langle U^{x>0}_g\rangle = \bra{\sigma_\mathrm{L}}\sigma^g_\mathrm{R}\rangle = \mathrm{Tr}\left[(\tau_g)^{L}\right],
\end{equation}
which is an overlap between two boundary MPSs along cylinder circumference at $x=0$ as shown in Fig.~\ref{fig:partial_sym_operation_without_W}(a), where we defined the tensor 
\begin{equation}\label{eq:tau_g_definition}
    (\tau_g)_{a,b,c,d} := \sum_{\alpha,\beta=1}^{D}(M_\mathrm{L})^{\alpha,\beta}_{a,b}(M^g_\mathrm{R})^{\alpha,\beta}_{c,d},
\end{equation}
and reshaped it to a matrix $(\tau_g)_{(ac),(bd)}$ [see Fig.~\ref{fig:partial_sym_operation_without_W}(b)]. The matrix $\tau_g$ plays the role of a transfer matrix when overlapping boundary MPSs.

\subsection{Main result}\label{subsec:main_claim}

With $\tau_g$ defined, we can state our main result:
for $G=\mathbb{Z}_2$, there exists an anti-linear symmetry $\mathcal{V} = (V_{\mathrm{L}}^{\mathrm{H}}\otimes V_{g})\mathcal{K}$ ($\mathcal{K}$ is complex conjugation) of $\tau_g$:
\begin{equation}\label{eq:tau_g_Z_2_case_constraint1}
\tau_g = \mathcal{V} \tau_g \mathcal{V}^{-1},
\end{equation}
where $V_{\mathrm{L}}^{\mathrm{H}}$ and $V_g$ act on the virtual spaces of $M_{\mathrm{L}}$ and $M^{g}_{\mathrm{R}}$, respectively. Moreover,
\begin{equation}\label{eq:tau_g_Z_2_case_constraint2}
\mathcal{V}^2  = (V_{\mathrm{L}}^{\mathrm{H}}V_{\mathrm{L}}^{\mathrm{H}*})
\otimes (V_g V_g^*) = \mathbb 1 \otimes \omega(g,g,g) \mathbb1,
\end{equation}
where $\omega$ represents the class $[\omega]\in\mathcal{H}^3(\mathbb{Z}_2,U(1))$. 
The proof of Eq.~\eqref{eq:tau_g_Z_2_case_constraint1} and Eq.~\eqref{eq:tau_g_Z_2_case_constraint2} is given in Sec.~\ref{subsec:tau_g_degeneracy_proof}.
Furthermore, we provide a procedure to extract $V_g$ numerically for generic states and use it to determine $\omega(g,g,g)$ directly\footnote{While it appears that this equation probes the raw $3$-cocycle instead of the class, in the specific case of the $\mathbb{Z}_2$ SPT, the true invariant (reviewed in Appendix~\ref{appendix:TQFT}) is equivalent to $\omega(g,g,g)$. When this cocycle element is mentioned in the text, we have this distinction in mind.}.

The anti-linear symmetry enforces that all non-zero eigenvalues of $\tau_g$  come in pairs $\{\lambda, \lambda^*\}$, regardless of $\omega(g,g,g)$. 
However, $\omega(g,g,g)$ determines whether nondegenerate real eigenvalues can exist:
\begin{enumerate}
    \item If $\omega(g,g,g) = -1$, real nonzero eigenvalues form doubly-degenerate Kramers pairs, and will not be pinned to the real axis under generic deformations. 
    \item If $\omega(g,g,g) = 1$, no such Kramers-type constraint arises; at the fixed point of the trivial paramagnet [see Sec.~\ref{sec:example}], $\tau_g = \tau$ has a gapped spectrum and a non-degenerate real leading eigenvalue. By continuity, $\tau_g$ will retain the unique leading eigenvalue under perturbations.
\end{enumerate}
These two cases respectively correspond to the nontrivial/trivial $\mathbb{Z}_2$ SPT phase\footnote{For larger symmetry groups, more involved diagnostics are required, as for string-order parameters in 1D~\cite{PhysRevB.86.125441, PhysRevLett.109.050402}. For example, one can consider line-like operator insertions, in direct analogy to $\obs_{A/B}$ in the 1D MPS case [see Eq.~\eqref{eq:string_order_parameter}].}; we will see an example of this in Sec.~\ref{sec:example}.
Let us connect this claim to observables. Let the eigenvalues (sorted in magnitude in descending order) of $\tau_g$ be $\lambda_{a,g}$, $a=1,\ldots, \chi^2$. If $\beta_g$ denotes the degeneracy of the leading $\tau_g$ eigenvalue in magnitude, we have
\begin{equation}
    \langle U^{x>0}_g \rangle = \mathrm{Tr}[\tau_g^L] \xrightarrow {L \rightarrow \infty} \sum_{a=1}^{\beta_g}\lambda_{a,g}^L.
\end{equation}
 Writing these leading eigenvalues as $\lambda_{a,g} = e^{-\alpha_g} e^{\im\theta_{a,g}}$, where $e^{-\alpha_g}$ denotes their common magnitude, we have
\begin{equation}\label{eq:partial_symmetry_explicit_scaling}
     \langle U^{x>0}_g \rangle \approx e^{-\alpha_g L}\sum_{a=1}^{\beta_g} e^{\im \theta_{a,g} L}.
 \end{equation}
We identify $f(L) = \sum_a e^{\im \theta_{a,g} L}$ here.
Thus, in the nontrivial SPT phase, we expect oscillations of this observable~\cite{PhysRevLett.133.106503, Xu_2026} encoded in $f(L)$. We will see this in more detail in Sec.~\ref{sec:example}.

\subsection{Practical recipe}
To obtain $\tau_g$ numerically, we require the translationally invariant form of leading eigenvectors of TMs as in Fig.~\ref{fig:partial_sym_operation_without_W}.
To obtain it efficiently, we use the corner transfer matrix renormalization group (CTMRG) method~\cite{Nishino_1996,PhysRevB.84.041108}; one can equivalently use boundary MPS methods~\cite{Nietner2020efficient}.
The algorithm takes the iPEPS tensors $A$ or directly double-layer tensors $T$, $T_g$ as input and outputs the environment tensors, including  $M_\mathrm{L/R}$ and $M^g_\mathrm{L/R}$, that approximate the semi-infinite tensor contractions. Specifically,  $M_\mathrm{L}$ and $M^g_\mathrm{R}$ give uniform MPS representations of the leading eigenvectors $\langle\sigma_{\mathrm{L}}|$ and $|\sigma^g_{\mathrm{R}}\rangle$ of the transfer matrices $\mathcal{T}$ and $\mathcal{T}_g$, as shown in Fig.~\ref{fig:transfer_matrices_2D}. $M_\mathrm{L}, M^g_\mathrm{R}$ are then contracted as in Eq.~\eqref{eq:tau_g_definition} [see also Fig.~\ref{fig:partial_sym_operation_without_W}(b)] to obtain $\tau_g$. Using the method detailed in Appendix~\ref{app:Get_Vg}, we extract the gauge matrix $V_g$ from $M_{\mathrm{R}}^g$ and determine $\omega(g,g,g)$ via Eq.~\eqref{eq:tau_g_Z_2_case_constraint2}.
The accuracy of the approximation is controlled by the environment bond dimension $\chi$ [see Fig.~\ref{fig:transfer_matrices_2D}(a)].
The complexity of the method, dominated by CTMRG, is $O(\chi^3D^6)$.

\subsection{$\tau_g$ and the $3$-cocycle of the SPT}\label{subsec:tau_g_degeneracy_proof}
\begin{figure}
    \centering    \includegraphics[width=0.85\linewidth]{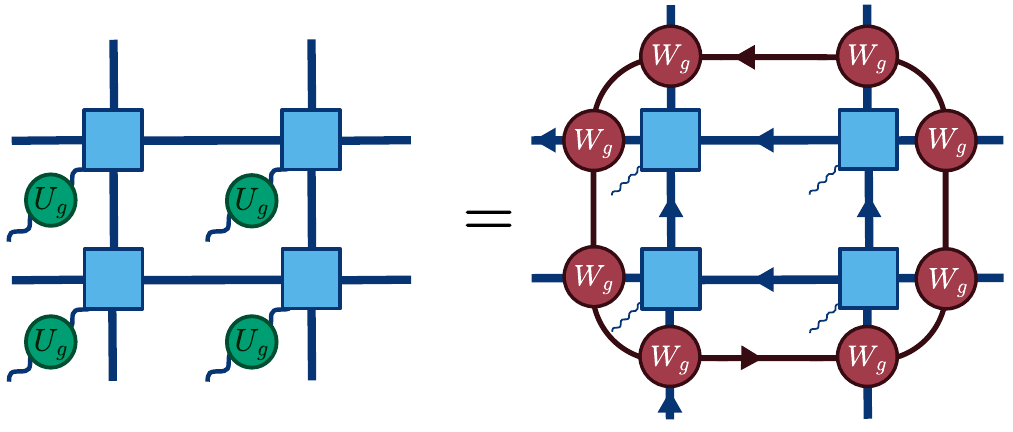}
    \caption{General definition of the pulling-through relation for arbitrary PEPS tensors, where $W_g$ is a MPO with some internal bond dimension (whose orientation matters~\cite{PhysRevB.94.205150}, shown by arrows). When contracting virtual legs of neighbor sites, these virtual MPOs cancel. In principle, one can obtain information about the SPT phase from $W_g$. For details, see Ref.~\onlinecite{PhysRevB.94.205150}.
    }    \label{fig:pulling_through_relation}
\end{figure}

We now derive the claim of Sec.~\ref{subsec:main_claim}.
For this, we first define virtual symmetry representations on PEPS symmetric under the global symmetry group $G$. In a similar vein to the 1D MPS case [Eq.~\eqref{eq:pulling_through_condition_1D}], previous works~\cite{PhysRevB.94.205150, 10.1063/1.5007017, PhysRevB.95.125107, RevModPhys.93.045003} established that the physical (on-site) symmetry $U_g$ acting on the PEPS tensor $A$ is equivalent to a symmetry action $\mathcal{W}_g$ on the virtual legs of the PEPS tensor, known as the pulling-through relation [see Fig.~\ref{fig:pulling_through_relation}]. Concretely, for any simply
connected region $\Sigma$, the blocked tensor $A_{\Sigma}$ obeys
\begin{equation}\label{eq:pull-through}
    U_g^{\Sigma} A_{\Sigma}
    = A_{\Sigma}\mathcal W_g^{\partial\Sigma}.
\end{equation}
$\mathcal W_g^{\partial\Sigma}$ is an MPO acting on the virtual boundary.
This relation is only true for a subset of PEPS, known as MPO-injective, semi-injective~\cite{10.1063/1.5007017} or $G$-injective~\cite{SCHUCH20102153}. In simple cases, such as in many renormalization group fixed points~\cite{PhysRevB.94.205150}, this action can be factorized on separate virtual legs as in Fig.~\ref{fig:pulling_through_relation}(a). Its most generic form, however, is a MPO acting on virtual legs, as in Fig.~\ref{fig:pulling_through_relation}(b). When contracting virtual legs of neighbor sites, such virtual symmetry MPOs cancel in the bulk, leaving non-trivial action only at the boundary. It has been shown that the information about the SPT phase (in other words, the $3$-cocycle class $[\omega]$) may be extracted from the pulled-through virtual symmetry MPO $\mathcal{W}_g$~\cite{PhysRevB.94.205150}, as we review in Appendix~\ref{appendix:MPO_review}. One method to extract SPT invariants would involve solving for $\mathcal{W}_g$ of the PEPS ground state with symmetry $U_g$. We discuss this in more detail in Appendix~\ref{appendix:variational_W}. 

\begin{figure}
    \centering
    \includegraphics[width=\linewidth]{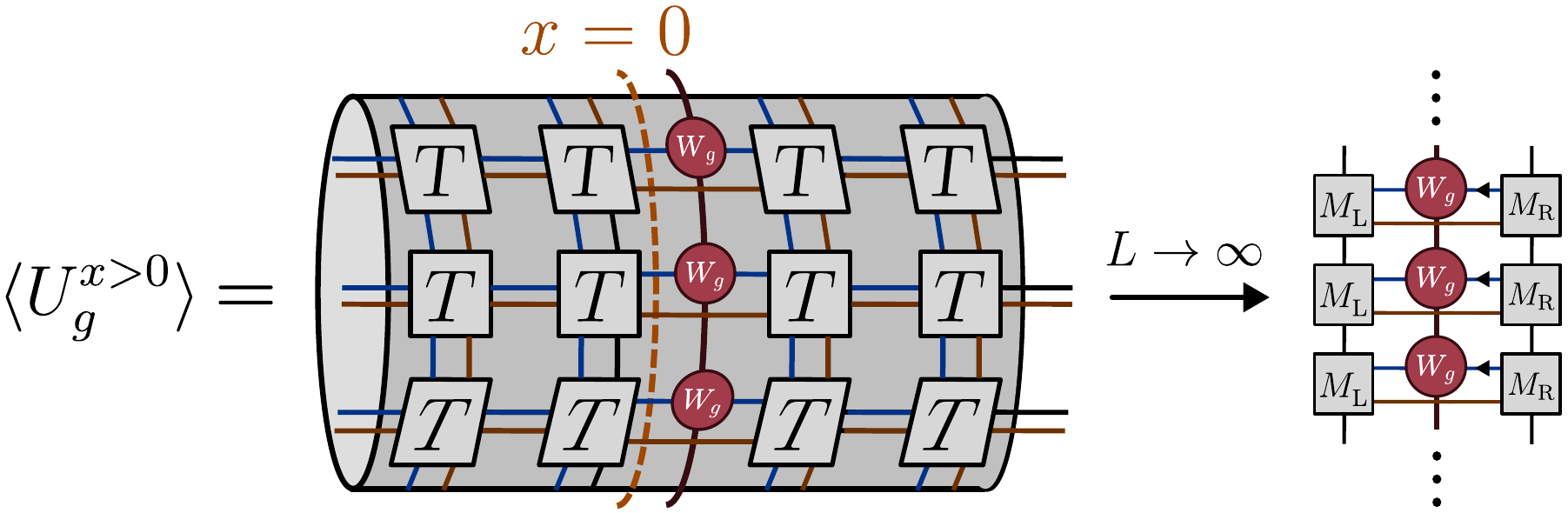}
    \caption{Interpretation of $\langle U^{x>0}_g \rangle$ with the virtual symmetry operation $W_g$ defined in Fig.~\ref{fig:pulling_through_relation}. Acting with $U_g$ on every site in region $R$ pulls through to virtual symmetries $W_g$ on all virtual legs, which cancel in the bulk~\cite{PhysRevB.94.205150} to leave only a string of $W_g$ operators at the boundary. It is understood that one should repeat the transfer matrices infinitely in all directions of the plane. In the thermodynamic limit, this contraction then corresponds to the overlap of $\sigma_\mathrm{L}$ and $\sigma_\mathrm{R}$ with the string $\mathcal{W}_g$ of virtual symmetry operators in between. This is a 2D analog of Fig.~\ref{fig:transfer_matrices_1D}(d) for SOPs.}
    \label{fig:partial_sym_operation_with_W}
\end{figure}

The virtual MPO $\mathcal{W}_g$ allows for a compact expression of $\langle U^{x>0}_g \rangle$. For clarity, we reshape the boundary MPS $|\sigma_{\mathrm{L/R}}\rangle$ into an MPO $\sigma_\mathrm{L/R}$, whose row and column indices correspond to the bra- and ket-layer indices of $|\sigma_{\mathrm{L/R}}\rangle$, respectively.
Pulling $U_g$ through the ket layer and representing the resulting contraction of $\langle \psi | \psi \rangle$ by the left boundary MPS $\sigma_{\mathrm{L}}$ and the right boundary MPS $\sigma_{\mathrm{R}}$, we have\footnote{Here the conventions are: (1) the row and column indices of $\sigma_{\mathrm{L/R}}$ correspond to the virtual indices of bra and ket iPEPS layers, respectively; (2) the row and column indices of $\mathcal{W}_g$ correspond to the in and out indices of the MPO, respectively.}
\begin{equation}\label{eq:partial_symmetry_trace}
    \langle U^{x>0}_g \rangle = \mathrm{Tr}(\sigma^T_\mathrm{L}\sigma_\mathrm{R} \mathcal{W}_g),
\end{equation}
which can be regarded as a product of MPOs. For reference, the ground state norm $\bra{\psi}\psi\rangle$ is given by $\mathrm{Tr}(\sigma^T_\mathrm{L}\sigma_\mathrm{R})$. Comparing to the 1D SPT detection via SOPs, Eq.~\eqref{eq:partial_symmetry_trace} is analogous to the 0D contraction of Fig.~\ref{fig:transfer_matrices_1D}(d) with $\obs_A = \obs_B = \id$. In addition, canonicalization of injective MPS in the 1D case allows one to choose $\sigma_\mathrm{R}=\id$, which is not true for 2D PEPS anymore. Thus, $\langle U^{x>0}_g \rangle$, or in general $\langle U^{\Sigma}_g \rangle$, is sensitive to the presence of the 1D line-like symmetry defect on the boundary $\partial\Sigma$ formed by $\mathcal{W}_g$ in Fig.~\ref{fig:partial_sym_operation_with_W}.

The partial symmetry $U_g^{x>0}$ can be either pulled through the ket or the bra layer, which implies a weak symmetry of $\sigma_{\mathrm{R}}$. Concretely, pulling $U_{g}^{x>0}$ through the bra layer results in the MPO $\mathcal{W}_{g^{-1}}^\dagger$, and we have
\begin{equation}\label{eq:weak_symm_mpo}
    \sigma^g_\mathrm{R} = \sigma_{\mathrm{R}} \mathcal{W}_g  = \mathcal{W}_{g^{-1}}^\dagger \sigma_{\mathrm{R}} = \mathcal{W}_g \sigma_{\mathrm{R}},
\end{equation}
where we have assumed that $\mathcal{W}_g$ is unitary in the last equality.
A similar weak symmetry exists for $\sigma_{\mathrm{L}}$.
In addition, both $\sigma_{\mathrm{L}}$ and $\sigma_{\mathrm{R}}$ are Hermitian, i.e., invariant under complex conjugation and permutation of the bra-layer and ket-layer legs (orange and blue legs respectively) in Fig.~\ref{fig:transfer_matrices_2D}, 
\begin{equation}\label{eq:Hermitian_MPO}
    \sigma_{\mathrm{L}}^\dagger = \sigma_{\mathrm{L}}, \quad 
    \sigma_{\mathrm{R}}^\dagger = \sigma_{\mathrm{R}}.
\end{equation}
Note that Eq.~\eqref{eq:weak_symm_mpo} and Eq.~\eqref{eq:Hermitian_MPO} are MPO equations, which hold for any length $L$.
Specializing to $G=\mathbb{Z}_2$, with $g$ the non-trivial element, we have
\begin{equation}\label{eq:hermitian_sigma_W}
    (\sigma_{\mathrm{R}} \mathcal{W}_g)^\dagger = \mathcal{W}_{g^{-1}} \sigma_{\mathrm{R}}
     = \sigma_{\mathrm{R}} \mathcal{W}_{g^{-1}} = \sigma_{\mathrm{R}} \mathcal{W}_g,
\end{equation}
i.e., $\sigma_{\mathrm{R}}\mathcal{W}_g$ is Hermitian\footnote{We confirm this numerically, where $(\sigma_{\mathrm{R}} W_g)^\dagger = e^{i\theta} \sigma_{\mathrm{R}}W_g$ due to the $U(1)$ phase freedom of CTM environments.}.

These relations have corresponding local implications on the MPO tensors.
We assume that the boundary MPS tensors $M_{\mathrm{L/R}}$ of $\sigma_{\mathrm{L/R}}$ are injective.
Equation~\eqref{eq:Hermitian_MPO} implies~\cite{PhysRevLett.100.167202}
\begin{equation}\label{eq:env_Hermitian}
\includegraphics[height=0.2\textheight]
    {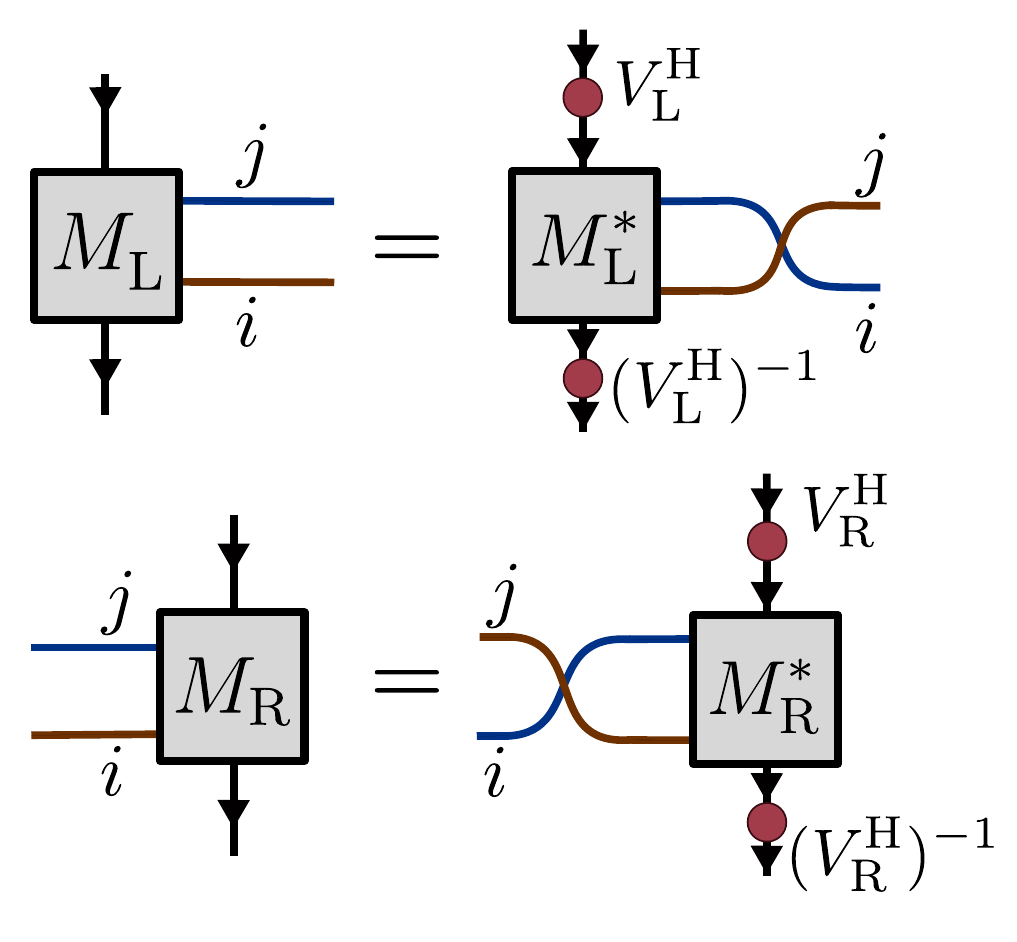} \quad \raisebox{0.05\textheight}{,} \\
\end{equation}
where $V_{\mathrm{L}/\mathrm{R}}^{\mathrm{H}}$ are virtual gauge transformations. As explicit equations, this gives
\begin{equation}\label{eq:env_Hermitian_eq}
\begin{aligned}
    M_{\mathrm{L}}^{ij} &= V_{\mathrm{L}}^{\mathrm{H}}M_{\mathrm{L}}^{*ji}
    (V_{\mathrm{L}}^{\mathrm{H}})^{-1}, \\
    M_{\mathrm{R}}^{ij} &= V_{\mathrm{R}}^{\mathrm{H}}M_{\mathrm{R}}^{*ji}
    (V_{\mathrm{R}}^{\mathrm{H}})^{-1},
\end{aligned}
\end{equation}
where $i,j$ index virtual states in bra (orange) and ket (blue) iPEPS layers.
Moreover, as shown in Appendix~\ref{app:VL_VL*}, the gauge matrices satisfy
\begin{equation}\label{eq:Hermitian_gauge}
V^{\mathrm{H}}_{\mathrm{L}} V^{\mathrm{H}*}_{\mathrm{L}} = \mathbb 1, \quad V_{\mathrm{R}}^{\mathrm{H}} V_{\mathrm{R}}^{\mathrm{H}*} = \mathbb 1.
\end{equation}

For the $\mathbb{Z}_2$ symmetry group, we have the additional Hermiticity condition [Eq.~\eqref{eq:hermitian_sigma_W}]. Since $M_{\mathrm{R}}$ is injective, one can show that $\sigma_{\mathrm{R}} \mathcal{W}_g$ admits an injective tensor representation [see Appendix~\ref{app:injective_sigma_R_Wg}], which we denote by $M_{\mathrm{R}}^g$. Note the raw tensor $M_{\mathrm{R}}W_g$ is not necessarily injective. Then we have a similar relation between $M_{\mathrm{R}}^g$ and $M_{\mathrm{R}}^{g*}$
\begin{equation}\label{eq:sigma_R_W_g_mpo}
\includegraphics[height=0.09\textheight]
    {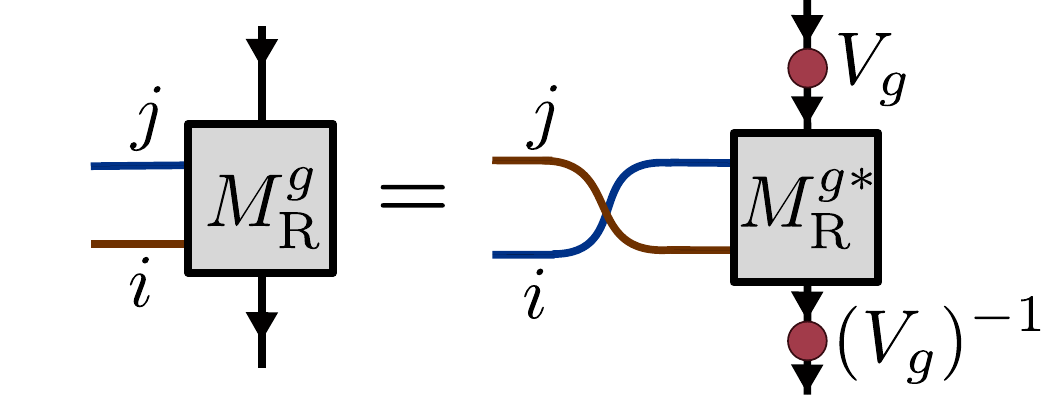} 
\end{equation}
with gauge matrix $V_g$. 
In terms of an equation, we have
\begin{equation}\label{eq:sigma_R_W_g_mpo_eq}
    (M_{\mathrm R}^{g})^{ij}
    =
    V_g (M_{\mathrm R}^{g})^{*ji}V_g^{-1}.
\end{equation}
Crucially, we prove in Appendix~\ref{app:VgR_VgR*} that 
\begin{equation}\label{eq:Vg_Vg*}
    V_g V_g^* = \omega(g,g,g) \mathbb{1},
\end{equation}
where $\omega(g,g,g)$ is the 3-cocycle labeling the $\mathbb{Z}_2$ SPTs.
In Appendix~\ref{app:Get_Vg}, we describe how to extract $V_g$ by considering the transfer matrices made of $M_{\mathrm{R}}^g$. Together with Eq.~\eqref{eq:Vg_Vg*}, we are able to calculate $\omega(g,g,g)$ directly. The same method is also used to find $V_{\mathrm{L/R}}^{\mathrm{H}}$ to numerically verify Eq.~\eqref{eq:Hermitian_gauge}.

Now, from Eq.~\eqref{eq:partial_symmetry_trace}, we have
\begin{equation}
\includegraphics[height=0.095\textheight]{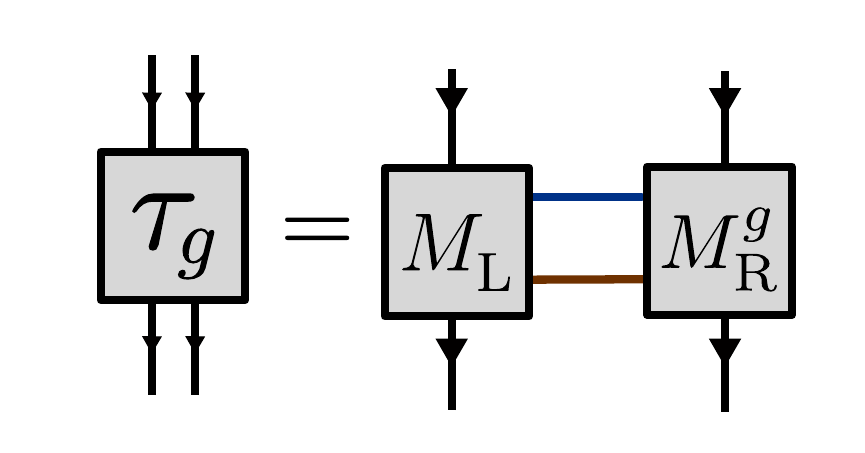}\raisebox{0.04\textheight}{.}
\end{equation}
This transfer matrix satisfies
\begin{equation}
    \tau_g = (V_{\mathrm{L}}^{\mathrm{H}} \otimes V_g) \tau_g^* (V_{\mathrm{L}}^{\mathrm{H}}\otimes V_g)^{-1}, 
\end{equation}
giving the anti-unitary $\mathcal{V} = (V_{\mathrm{L}}^{\mathrm{H}} \otimes V_g)\mathcal{K}$. The symmetry squares to
\begin{equation}\label{eq:main result Z2}
\begin{aligned}
    \mathcal{V}^2 &= (V_{\mathrm{L}}^{\mathrm{H}} \otimes V_g) (V_{\mathrm{L}}^{\mathrm{H}*} \otimes V_g^*) \\
    &= V_{\mathrm{L}}^{\mathrm{H}} V_{\mathrm{L}}^{\mathrm{H}*} \otimes V_g V_g^* = \omega(g,g,g) \mathbb{1}.
\end{aligned}
\end{equation}
This proves our claim in Sec.~\ref{subsec:main_claim}. For $\mathbb{Z}_2$ SPT phases, the minimal cocycle representing the non-trivial phase has $\omega(g,g,g)=-1$, with other elements being unity~\cite{CZX_model_paper, Chen_Gu_Wen_SPT_2013, PhysRevB.94.205150}. 

Equation~\eqref{eq:main result Z2} implies one of two options for $\tau_g$ if $\omega(g,g,g)=-1$: (i) $\tau_g=0$, (ii) the real eigenvalues of $\tau_g$ are doubly-degenerate and the complex eigenvalues of $\tau_g$ come in magnitude-degenerate pairs. In the trivial SPT phase, we have $\omega(g,g,g)=+1$ and $\mathcal{W}_g=\id$ (at least at the fixed point), i.e., $\tau_g=\tau$ up to a gauge. In this case, the leading eigenvalue of $\tau_g$ is real without any degeneracy.

The above argument correctly probes the partition function for so-called type I SPTs~\cite{PhysRevB.96.195101, PhysRevB.91.195134}, of which the CZX model is one example.
It might come as a surprise that the diagnostic we present appears local, in contradiction with conventional wisdom that non-local order parameters are required to probe topological phases. This is resolved by the fact that the MPSs we are using represent half-infinite contractions of TMs, which are clearly non-local, in the same vein as the simplification of the string-order parameter to a $0$D contraction in Sec.~\ref{sec:1D_section}.

\section{Example: the CZX model}\label{sec:example}
\begin{figure}
    \centering
    \includegraphics[width=\linewidth]{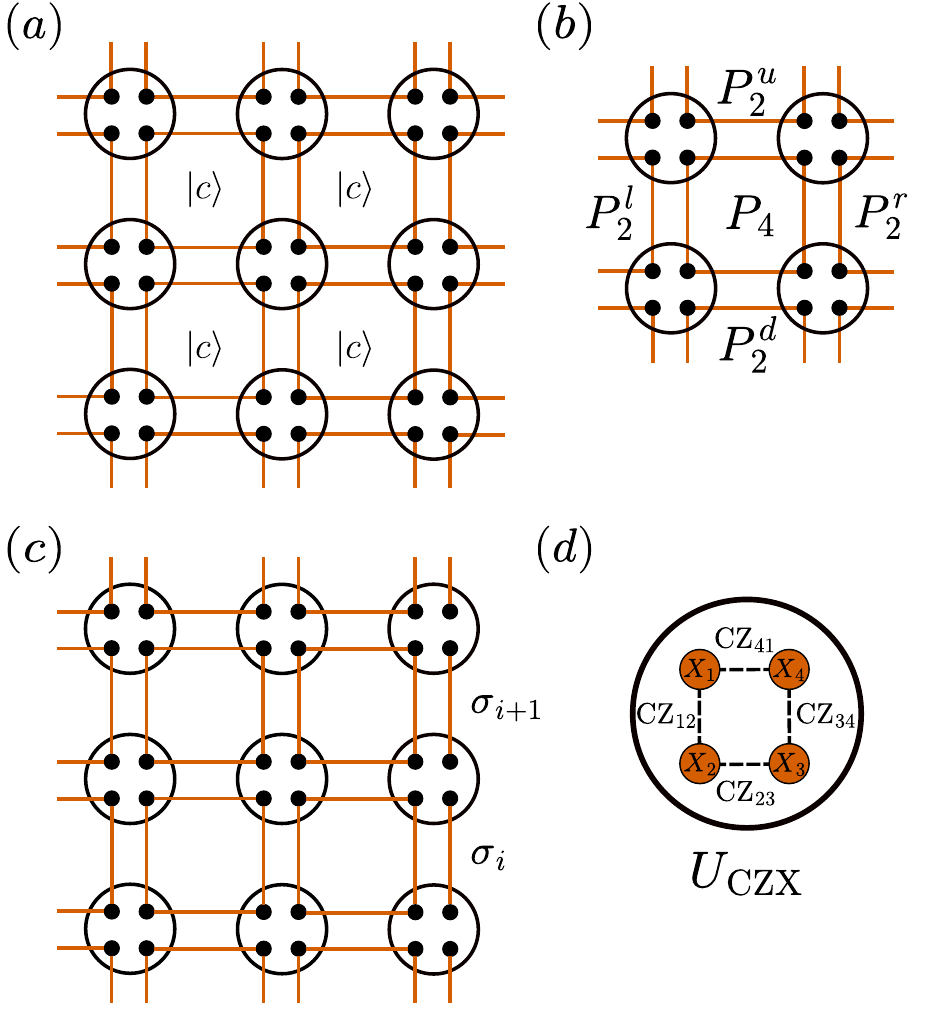}
    \caption{(a) The ground state of the CZX Hamiltonian, where $\ket{c} = \frac{1}{\sqrt{2}}(\ket{0000}+\ket{1111})$ is a four-spin maximally entangled state on each plaquette. (b) One term in the CZX Hamiltonian [Eq.~\eqref{eq:CZX_hamiltonian}], involving a plaquette term $P_4$ as well as terms $P_2$ in all four neighbor half-plaquettes. (c) Along an open boundary, the spins in  CZX ground state are not entirely constrained, and can be grouped into effective qubits $\sigma_i$~\cite{CZX_model_paper, wang_wen_witten_2018}. (d) The on-site $\mathbb{Z}_2$ symmetry operation $U_\CZX$ [Eq.~\eqref{eq:CZX_symmetry}] which protects the SPT phase.}
    \label{fig:CZX_model}
\end{figure}

\begin{figure}
    \centering
    \includegraphics[width=\linewidth]{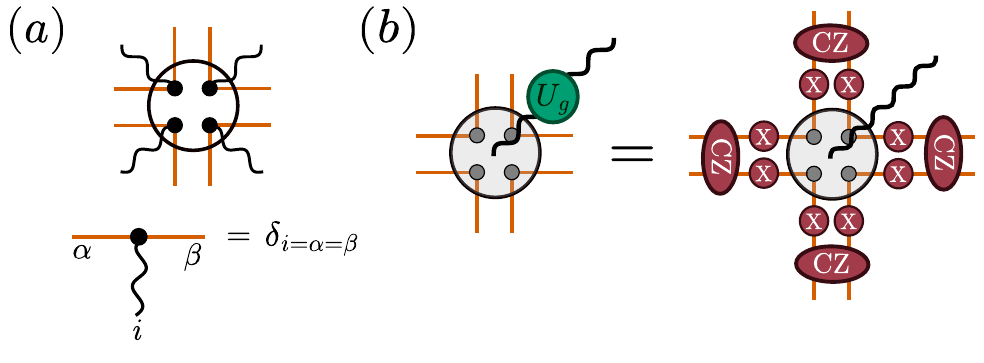}
    \caption{(a) The PEPS tensor for the CZX fixed point. It can be constructed as a tensor product of four MPS tensors for cat states $A^{i}_{\alpha \beta} =\delta_{i=\alpha=\beta}$. When contracting multiple PEPS tensors, the corresponding state on plaquettes will indeed be a $4$-particle cat state, as in Fig.~\ref{fig:CZX_model}(a). (b) The physical symmetry action $U_g$ corresponds to actions on the virtual legs as shown. This can be proven straight from the tensor product of the four cat state MPSs. This is the pulling-through relation of Fig.~\ref{fig:pulling_through_relation}(a). The order with which we ``read" the actions of $X$ and $\CZ$ on virtual legs matters: these actions do cancel in the bulk, and here $W_\CZX = (X\otimes X)\CZ$ on top and left legs, and $W^\dagger_\CZX = \CZ(X\otimes X)$ on bottom and right legs.}
    \label{fig:CZX_model_PEPS}
\end{figure}

Let us now apply the method to the well-studied example of a 2D SPT known as the CZX model~\cite{CZX_model_paper, Chen_Gu_Wen_SPT_2013, wang_wen_witten_2018}, which is protected by an on-site $\mathbb{Z}_2$ unitary symmetry.
Consider a square lattice shown in Fig.~\ref{fig:CZX_model}(a), where each site $i$ hosts a spin-1/2 physical  Hilbert space, with basis labelled $\{\ket{0},\ket{1}\}$ for spin-down and spin-up, respectively.
Quadruplets of spins are grouped into sites, forming a $16$-dimensional single-site Hilbert space. Plaquettes, on the other hand, comprise the 4 corner spins of 4 different sites. On each plaquette, we define
\begin{equation}
    P_4 = \ket{0000}\bra{1111} + \ket{1111}\bra{0000},
\end{equation}
which interchanges the two symmetry-breaking states. Similarly, on spins from neighboring sites we define
\begin{equation}
    P_2 = \ket{00}\bra{00} + \ket{11}\bra{11}.
\end{equation}
The CZX Hamiltonian is then defined as a sum over plaquettes $p$ via
\begin{equation}\label{eq:CZX_hamiltonian}
    H_\CZX = -\sum_p P_4 \otimes P_2^u \otimes P^r_2 \otimes P^d_2 \otimes P^l_2, 
\end{equation}
where we introduced labels for the four $P_2$ operators that neighbor a given plaquette, as shown in Fig.~\ref{fig:CZX_model}(b). This Hamiltonian is purely interacting: band theory and single-particle invariants~\cite{doi:10.1126/science.1227224, PhysRevB.76.045302} do not distinguish this model from a trivial paramagnet [which we define in Eq.~\eqref{eq:PM_hamiltonian}].
The CZX ground state is a tensor product of maximally entangled states on plaquettes
\begin{equation}\label{eq:CZX_model_GS}
    \ket{\CZX} = \bigotimes_{p} \frac{1}{\sqrt{2}}(\ket{0000} + \ket{1111}).
\end{equation}
One can construct the PEPS tensor representing the CZX model ground state, as shown in Fig.~\ref{fig:CZX_model_PEPS}(a). We now define the on-site symmetry $U_\CZX = U_\mathrm{X} U_\CZ$, whose single-site action is
\begin{equation}\label{eq:CZX_symmetry}
\begin{aligned}
    U_\mathrm{X} &= X_1X_2X_3X_4,\\
    U_\CZ &= \CZ_{12}\CZ_{23}\CZ_{34}\CZ_{41},
\end{aligned}
\end{equation}
where we labelled a site's four spins as in Fig.~\ref{fig:CZX_model}(d), and used the notation $X,Y,Z$ for Pauli matrices and $\CZ$ for a controlled-Z quantum gate, which acts as $\mathrm{diag}(1,1,1,-1)$ on pairs of spins.
The Hamiltonian $H_{\CZX}$ is invariant under the action of $\UCZX$. The ground state is also symmetric under $\UCZX$ with eigenvalue $+1$.
By virtue of $\UCZX$ being a symmetry of this ground state, the action of this symmetry on a subregion $\Sigma$ should only have a non-trivial action near the boundary $\partial \Sigma$ of that region. It was shown that this low-energy effective action could in fact not be factorized on sites~\cite{CZX_model_paper}, representing a quantum anomaly at the lattice level. Under open boundary conditions, the edge degrees of freedom are only partially constrained, giving degenerate gapless edge states which cannot be gapped out as long as $\UCZX$ is enforced. When imposing the symmetry $U_\CZX$, there are then two possible phases~\cite{CZX_model_paper} on this lattice, the non-trivial one being represented by the CZX model. A representative of the trivial phase can be constructed via the Hamiltonian
\begin{equation}\label{eq:PM_hamiltonian}
   \HPM = -\sum_{\text{sites}} \sum_{i=1}^4 X_{i}(\mathbb{1} + Z_{i-1} Z_{i+1}),
\end{equation}
whose unique ground state, which is symmetric under $\UCZX$, is a product state\footnote{Its precise ground state (up to normalization) is a product of \begin{equation}\label{eq:PM_GS}
\begin{aligned}
\ket{\PM} = \sqrt{2}&\ket{0000} + \ket{1000} + \ket{0100} + 
\ket{0010} + \ket{0001} + \\ &\sqrt{2} \ket{1010} 
+ (0 \leftrightarrow 1).
\end{aligned}
\end{equation} Curiously, this Hamiltonian is precisely the one used to model the \emph{boundary} of the CZX model in Ref.~\onlinecite{PhysRevB.86.235135}.}. This ground state cannot be distinguished from the CZX ground state by its (unit) eigenvalue under $U_\CZX$; this is true regardless of system size. The pulled-through virtual symmetry operator $W_g$ (for $g=\CZX$) for the CZX fixed point is shown in Fig.~\ref{fig:CZX_model_PEPS}(b). In this case, $\mathcal{W}_g$ is just a matrix on each pair of virtual legs, and equals $\mathrm{CZ}(X \otimes X)$.
By contrast, the trivial paramagnet fixed point has a PEPS tensor that factorizes on each site, meaning the virtual legs are trivial. We simply have $\mathcal{W}_g=1$ in that case.

The $3$-cocycle class $[\omega] \in \mathcal{H}^3(G,U(1))$ that characterizes the CZX phase is distinct from that of the trivial phase. The CZX model cannot be adiabatically connected to the paramagnet without a topological phase transition.
As in the 1D case [see Sec.~\ref{sec:1D_section}], the entanglement spectrum provides a ``smoking gun" for the CZX phase~\cite{pollmann_entanglement_1d_2010, turner_pollmann_1d_top_fermions_entanglement, entang_spec_boundary_theories_PEPS_2011, PhysRevB.107.115113}.
In the non-trivial SPT phase, the entanglement spectrum displays a degeneracy that can be seen as coming from gapless edge states specified by effective boundary spins $\sigma_i, i = 1, \ldots, L$, that live on half-plaquettes~\cite{CZX_model_paper, wang_wen_witten_2018}, as in Fig.~\ref{fig:CZX_model}(c). The number of such degenerate levels scales exponentially with the size of the cylinder boundary as $2^L$. 
\begin{figure*}
    \centering
    \includegraphics[width=\linewidth]{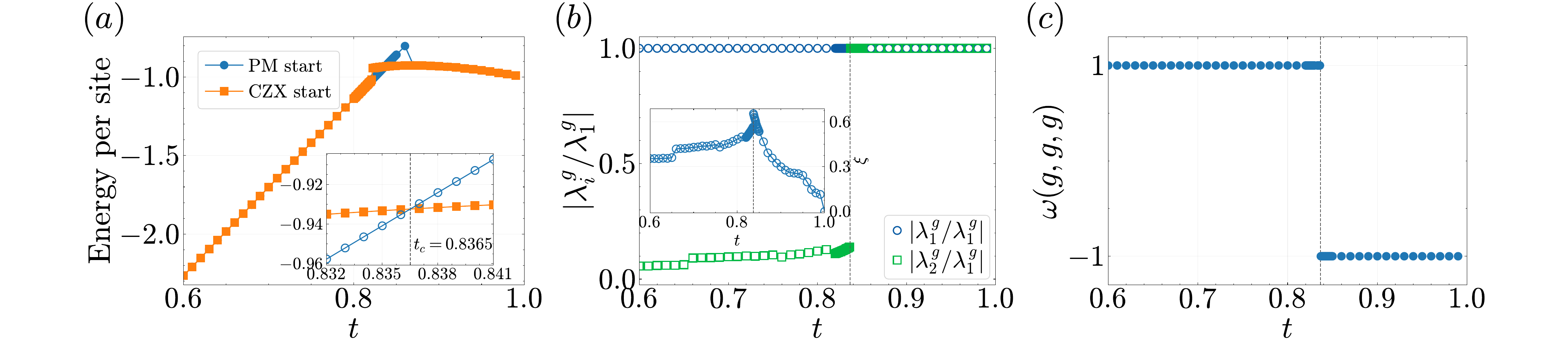}
    \caption{Results for the interpolation in Eq.~\eqref{eq:hamiltonian_interpolation}. (a) Variational ground-state energy throughout the interpolation, obtained by optimizing iPEPS with $D=5$ and $\chi_{\mathrm{opt}}=40$. 
    We conduct two optimization sweeps: one starts from the trivial paramagnetic fixed-point tensor at $t=0$ and increases $t$ (PM start), while the other starts from the CZX fixed-point tensor at $t=1$ and decreases $t$ (CZX start). In each sweep, each optimization is initialized with a slightly perturbed version of the previously converged state.
    The level crossing and the hysteresis are strong evidence for a first-order phase transition. 
    (b) Magnitudes of the two leading eigenvalues $\lambda_i^g$ of $\tau_g$ across the interpolation. For each $t$, we analyze the lower-energy state obtained from the two sweeps. The SPT phase exhibits the expected double degeneracy. We also observe that the magnitude $|\lambda^g_i| \rightarrow 0$ as $t \rightarrow 1$. This is expected as $\tau_g = 0$ for the CZX fixed point; for a proof, see Appendix~\ref{appendix:CZX_fixed_point}. The inset shows the correlation length $\xi$, calculated from the leading eigenvalues of $\tau$.
    (c) The $\mathbb{Z}_2$ SPT invariant $\omega(g,g,g)$, extracted numerically from the product $V_gV_g^*$ in Eq.~\eqref{eq:Vg_Vg*}. For the analysis in (b) and (c), we use $\chi_{\mathrm{eval}}=80$, and they remain the same when environment tensors with larger $\chi$ are used.
    }
\label{fig:final_phase_diagram}
\end{figure*}

To demonstrate the method, we consider a one-parameter interpolation between $H_{\CZX}$ and $H_\PM$ 
\begin{equation}\label{eq:hamiltonian_interpolation}
    H(t) = (1-t)H_{\PM} + t H_{\CZX},
\end{equation}
where $t\in [0,1]$. We variationally minimize the energy of iPEPS via gradient-based optimization~\cite{hasik2020peps}. We set the iPEPS bond dimension to $D=5$ and the environment bond dimension $\chi_{\mathrm{opt}}=40$.
To determine the phase transition, we conduct two optimization sweeps in opposite directions. 
The first starts from the trivial paramagnetic fixed-point state at $t=0$ and gradually increases $t$, initializing each optimization with a slightly perturbed version of the previously converged state. The second starts from the CZX fixed-point state at $t=1$ and follows the same procedure while decreasing $t$.
As shown in Fig.~\ref{fig:final_phase_diagram}(a), the resulting energies show a level crossing at $t_c\approx 0.837$, indicating a first-order phase transition. The correlation length $\xi$ also remains finite throughout the interpolation, as shown in the inset of Fig.~\ref{fig:final_phase_diagram}(b).

To investigate the SPT nature, we calculate the leading eigenvalues of $\tau_g$, as shown in Fig.~\ref{fig:final_phase_diagram}(b). We observe that the spectrum of $\tau_g$ is degenerate in magnitude for $t > t_c$, in contrast to the trivial phase. Consequently, we expect a nontrivial scaling function $f(L)$ in $\langle U^{x>0}_g \rangle$. In particular, the leading eigenvalues come in opposite pairs here, implying
\begin{equation}
    f(L) = 1+(-1)^L,
\end{equation}
up to an overall $U(1)$ phase. 

Both numerically and analytically, $\tau_g$ is exactly $0$ at the CZX fixed point. In Appendix~\ref{appendix:CZX_fixed_point}, we prove this fact for the CZX model with the explicit PEPS representation. An analogous argument proves that $\tau_g=0$ for the Levin-Gu wavefunction~\cite{PhysRevB.86.115109, PhysRevX.5.041034}, which is another version of the $\mathbb{Z}_2$ SPT. 
Note that we do not expect $f(L)$ to be a simple constant greater than $1$ for the following reason. From the perspective of the gauged anyon theory~\cite{PhysRevB.86.115109, PhysRevB.91.195134}, $\langle U^{x>0}_g \rangle$ is interpreted as the quantum dimension of the corresponding $g$-defect~\cite{PhysRevB.106.094415}. Yet gauging the CZX model gives the double-semion theory, which only contains Abelian anyons, i.e., their quantum dimension is $1$.

Furthermore, since we have obtained the edge tensor $M^g_{\mathrm{R}}$ from applying CTMRG to the symmetry-inserted double-layer tensor $\mathcal{T}_g$, we can directly determine the gauge matrix $V_g$ and, through Eq.~\eqref{eq:Vg_Vg*}, extract the SPT label $\omega(g,g,g)$. As shown in Fig.~\ref{fig:final_phase_diagram}(c), the computed value of $\omega(g,g,g)$ provides a clear diagnostic for distinguishing the two $\mathbb{Z}_2$-symmetric phases. In Appendix~\ref{appendix:CZX_fixed_point}, we explicitly obtain $V_g$ at the CZX fixed point ($t=1$) and extract $\omega(g,g,g)=-1$ analytically.

To further probe the robustness of the diagnostic, we also generated random on-site deformations $\epsilon \delta H$ to $H_\CZX$, where $\epsilon \in \mathbb{R}^+$ controls the perturbation strength. These Hamiltonian deformations are then Hermiticized and symmetrized with respect to $U_\CZX$. Starting from the CZX fixed point for which $\tau_g = 0$, we find that small enough $\epsilon$ does give the pairing of eigenvalues in $\tau_g$. For large enough $\epsilon$ we observe a single leading eigenvalue in $\tau_g$, indicating a phase transition to the trivial phase.

\section{Discussion}\label{sec:conclusion}
Using the language of PEPS, we have established a method to determine whether a physical symmetry operation $U_g$ protects a trivial/non-trivial $\mathbb{Z}_2$ SPT phase. The method does not require knowledge of the virtual symmetry operations on the PEPS tensor and  it efficiently extracts the SPT label in the thermodynamic limit. 

The idea to extract topological information from symmetry actions on subregions of a system was recently adapted to topological phases in open systems~\cite{PhysRevLett.133.106503, PhysRevB.111.205125, ljvt-w6hw}. In this spirit, one can interpret our method as effectively detecting the anomaly of the effective boundary symmetry on the mixed-state reduced density matrix at the boundary of the system~\cite{PhysRevB.94.205150, Sahinoglu2021}. More explicitly, consider the leading left/right eigenvectors $\sigma_{\mathrm{L}/\mathrm{R}}$ of $\mathcal{T}$ [see Fig.~\ref{fig:transfer_matrices_2D}], reshaped to matrices. From these we define the ``virtual" boundary density matrix
\begin{equation}\label{eq:virtual_RDM_definition}
    \sigma = \sqrt{\sigma^T_{\mathrm{L}}} \sigma_\mathrm{R} \sqrt{\sigma^T_\mathrm{L}}.
\end{equation}
The virtual boundary density matrix $\sigma$ is related to the physical reduced density matrix $\rho_{x\mathrm{>0}} := \mathrm{Tr}_{x\mathrm{<0}}(\rho)$, obtained by tracing out the left ($x<0$) part of the system [see Fig.~\ref{fig:partial_sym_operation_without_W}], by an isometry~\cite{PhysRevLett.111.090501, PhysRevB.107.115113} $Q$ via
\begin{equation}
    \rho_{x>0} = Q \sigma Q^\dagger,
\end{equation}
where $Q^\dagger Q = \id$. While this holds for infinite-length cylinders, this relation also exists for finite lengths, where $\sigma_{\mathrm{L},\mathrm{R}}$ are obtained from repeated contractions of TMs on a half-cylinder, together with edge tensors determining boundary conditions~\cite{entang_spec_boundary_theories_PEPS_2011}. The virtual matrix $\sigma$ has been interpreted as a one-dimensional density matrix at the boundary~\cite{PhysRevLett.111.090501, Xu_2026, PhysRevB.101.115143}, providing a tensor-network interpretation of the Li-Haldane conjecture~\cite{original_entang_spectrum_haldane}.
It seems natural to conjecture that $\sigma$ is classified by the average SPT phase framework of Ref.~\onlinecite{PhysRevX.15.021062}. 
The precise relation between the bulk pure SPT and this boundary mixed-state average SPT would be worth exploring in depth. 

Several other important questions are left open. The first obvious direction is to adapt such observables for more complicated symmetry groups, in direct analogy with 1D string-order selection rules~\cite{PhysRevB.86.125441}. It would also be worth exploring the applicability of this method to other representations of the ground state wavefunction, such as Monte Carlo simulations. In addition, the partial symmetry operations used here have been linked to path integrals on topological surfaces~\cite{PhysRevB.86.125441, PhysRevLett.118.216402}. It would be worth investigating how translational invariance allows one to further simplify such observables in the path integral formulation, to obtain the analog of the $\tau_g$ operator in this setting. Lastly, an in-depth study of the behavior of this operator at phase transitions could reveal information about the conformal field theory describing the system there~\cite{PhysRevLett.134.146503}. 

\begin{acknowledgments} 
We thank Frank Pollmann, Ramanjit Sohal, Anna Francuz, Derek K.K. Lee, Andras Molnar, Norbert Schuch, and Frank Schindler for insightful discussions. ChatGPT was used, under the authors' instruction, to coordinate numerical runs and help refine the proofs.

A.C. acknowledges support from Imperial College London via a President’s PhD Scholarship. 
H.C., J.H., and T.N. acknowledge support from the Swiss National Science Foundation through a Consolidator Grant (iTQC, TMCG-2213805) and a Quantum grant (20QU-1225225). 

\end{acknowledgments}

\bibliography{refs}

\appendix

\section{Topological quantum field theory interpretation}\label{appendix:TQFT}
As established in Sec.~\ref{subsec:tau_g_degeneracy_proof}, the order parameter $\langle U^{x>0}_g \rangle$ we use to probe the SPT phase relates to some aspect of the $3$-cocycle equivalence class describing the non-trivial SPT phase. More generally, there are known invariants, depending on the $3$-cocycle, that arise from partition functions on certain topological manifolds~\cite{PhysRevB.96.195101, mfwg-svcp} equipped with flat connections. Consider SPTs protected by symmetry group $G=\prod_{i=1}^{K}\mathbb{Z}_{N_i}$, for positive integers $N_i$ and $K$, where the generator of each $\mathbb{Z}_{N_i}$ subgroup is denoted by $e_i$. Type I SPTs~\cite{PhysRevB.91.195134}, which include the $K=1$ case, have corresponding partition functions
\begin{equation}
    Z_i = \prod_{n=1}^{N_i}\omega(e_i,ne_i,e_i).
\end{equation}
These are partition functions over the lens space $L(N_i;1)$ with generator $e_i$ placed on the single non-contractible loop of this manifold.
For the CZX model, $K=1$ and $N_{i=1}=2$, with generator $e_1=\CZX$. Then the single non-trivial partition function reduces to $\omega(g,g,g)\omega(g,\id,g)=-1\times 1 = -1$. This partition function is related to the topological spin $\Theta_i$ of the gauged $g$-anyon~\cite{PhysRevB.96.195101, PhysRevB.90.235113} via $Z_i=e^{2\pi i \Theta_i}$. Partition functions for type II and III SPTs exist as well~\cite{PhysRevB.96.195101}. As type-III SPTs can have non-Abelian gauged anyons, they can be probed by degeneracy (not just in magnitude) of the leading eigenvalues of $\tau_g$, corresponding to a quantum dimension greater than $1$ for the corresponding $g$-anyon.

\section{Review: MPO symmetries of two-dimensional SPTs}\label{appendix:MPO_review}

Two-dimensional SPT phases admit a natural
description in terms of PEPS with MPO symmetries.  Let $A$ be a PEPS tensor representing an
SPT phase protected by an on-site symmetry group $G$.  For every simply
connected region $\Sigma$, the blocked tensor $A_{\Sigma}$ obeys the
pull-through relation~\cite{PhysRevB.94.205150}
\begin{equation}\label{eq:pull-through-app}
    U_g^{\Sigma} A_{\Sigma}
    = A_{\Sigma}\mathcal W_g^{\partial\Sigma},
    \qquad g\in G .
\end{equation}
Here $A_{\Sigma}$ maps the virtual degrees of freedom on
$\partial\Sigma$ to the physical Hilbert space
$(\mathbb C^d)^{\otimes |\Sigma|}$,
$U_g^{\Sigma}=\bigotimes_{i\in\Sigma}U_{g,i}$ is the physical
on-site symmetry, and $\mathcal W_g^{\partial\Sigma}$ is an MPO acting on the virtual boundary.

Following the MPO-injective framework of Williamson et al.~\cite{PhysRevB.94.205150}, we
assume a compatible local MPO description of the virtual-support
projectors. Let $\mathcal H_{\partial\Sigma}$ denote the virtual
boundary Hilbert space, and define
\begin{equation}
    \mathcal S_{\Sigma}
    :=(\ker A_{\Sigma})^\perp,
    \qquad
    P_{\Sigma}
    :=A_{\Sigma}^{+}A_{\Sigma},
\end{equation}
where $A_{\Sigma}^{+}$ is the Moore--Penrose inverse.
Then $P_{\Sigma}=P_{\Sigma}^\dagger=P_{\Sigma}^2$
is the orthogonal projector onto $\mathcal S_{\Sigma}$,
$A_{\Sigma}P_{\Sigma}=A_{\Sigma}$, and
$A_{\Sigma}$ is injective on $\mathcal S_{\Sigma}$.
The nontrivial MPO assumption is that these projectors are
generated by fixed local MPO tensors, with bond dimension
independent of the boundary length and with the standard
compatibility conditions under concatenation of PEPS regions.

The pull-through equation uniquely determines the projected
virtual action $P_{\Sigma}\mathcal W_g^{\partial\Sigma}
P_{\Sigma}$, while components mapping into
$\ker A_{\Sigma}$ remain undetermined. We therefore replace
each virtual symmetry by its supported representative,
\begin{equation}\label{eq:supp_repr}
    \mathcal W_g^{\partial\Sigma}
    \longmapsto
    P_{\Sigma}\mathcal W_g^{\partial\Sigma}P_{\Sigma},
\end{equation}
which preserves the pull-through equation. We use the same
notation for this supported representative throughout.

Using $U_g^{\Sigma}U_h^{\Sigma}=U_{gh}^{\Sigma}$ in
Eq.~\eqref{eq:pull-through-app} and applying $A_{\Sigma}^{+}$ gives
\begin{equation}\label{eq:W_gh}
    P^{\partial\Sigma}
    \mathcal W_g^{\partial\Sigma}
    \mathcal W_h^{\partial\Sigma}
    =P^{\partial\Sigma}\mathcal W_{gh}^{\partial\Sigma}.
\end{equation}
Using the supported representative [Eq.~\eqref{eq:supp_repr}] and omitting the boundary label, this becomes
\begin{equation}
    \mathcal W_g\mathcal W_h=\mathcal W_{gh},
    \qquad \mathcal W_e=P.
\end{equation}
We further assume that $\mathcal{W}_g^{\partial\Sigma}$ is unitary on $\mathcal{S}_{\Sigma}$
\begin{equation}
    \mathcal{W}_g^\dagger = \mathcal{W}_{g^{-1}}, \, \mathcal{W}_g^\dagger \mathcal{W}_g = P.
\end{equation}
The fundamental theorem for MPOs provides a reduction tensor $R(g,h)$ and a
corresponding splitting tensor $S(g,h)$ such that
\begin{equation}\label{eq:reduction_split}
    \includegraphics[height=0.1\textheight]
        {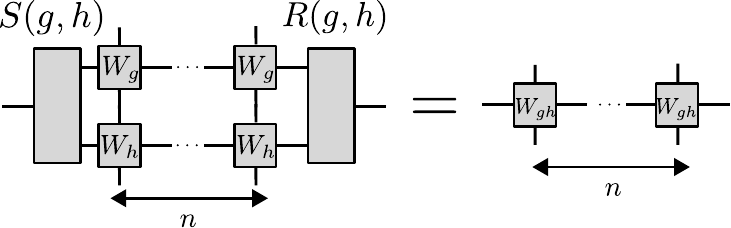}
    \quad \raisebox{0.045\textheight}{}
\end{equation}
for any length $n \geq 0$.

The 3-cocycle characterizing the SPT phase is obtained by reducing
$\mathcal W_g\mathcal W_h\mathcal W_k$ to $\mathcal W_{ghk}$ in the two possible orders:
\begin{equation}\label{eq:mpo_associativity_diagram}
    \includegraphics[height=0.08\textheight]
        {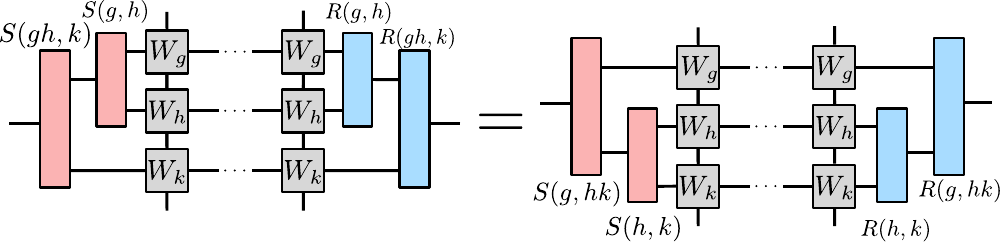}
    \quad \raisebox{0.035\textheight}{.}
\end{equation}
Because each $W_g$ is represented by a single injective block, the two reduction maps (blue) in Eq.~\eqref{eq:mpo_associativity_diagram} belong to the same
one-dimensional intertwiner space and must be proportional:
\begin{equation}\label{eq:F_associator}
\begin{aligned}
    R(gh,k)\bigl(R(g,h)\otimes\mathbb 1_k\bigr)
    =\, &\omega(g,h,k) R(g,hk) \\
      &\times \bigl(\mathbb 1_g\otimes R(h,k)\bigr),
\end{aligned}
\end{equation}
with $|\omega(g,h,k)|=1$.  The splitting tensors obey the inverse relation
\begin{equation}\label{eq:F_split_associator}
\begin{aligned}
    \bigl(S(g,h)\otimes\mathbb 1_k\bigr)S(gh,k)
    =\, &\omega(g,h,k)^{-1}  \bigl(\mathbb 1_g\otimes S(h,k)\bigr) \\
      & \times S(g,hk).
\end{aligned}
\end{equation}

Consistency of the reductions of four MPOs gives the pentagon equation
\begin{equation}\label{eq:three_cocycle}
    \omega(h,k,l)\,\omega(g,hk,l)\,\omega(g,h,k)
    =\omega(gh,k,l)\,\omega(g,h,kl),
\end{equation}
so $\omega$ is a group $3$-cocycle.  Finally, the redefinition
\begin{equation}
\begin{aligned}
    S(g,h) &\longmapsto\beta(g,h)^{-1}S(g,h), \\
    R(g,h)&\longmapsto\beta(g,h)R(g,h)
\end{aligned}
\end{equation}
changes $\omega$ by the coboundary
\begin{equation}
    \omega(g,h,k)\longmapsto
    \omega(g,h,k)
    \frac{\beta(gh,k)\beta(g,h)}
         {\beta(g,hk)\beta(h,k)}.
\end{equation}
Consequently, the gauge-invariant datum carried by the MPO reduction is the cohomology class $[\omega]\in H^3(G,U(1))$.

\section{Variational solutions of virtual symmetries}\label{appendix:variational_W}
\begin{figure}
    \centering
    \includegraphics[width=0.95\linewidth]{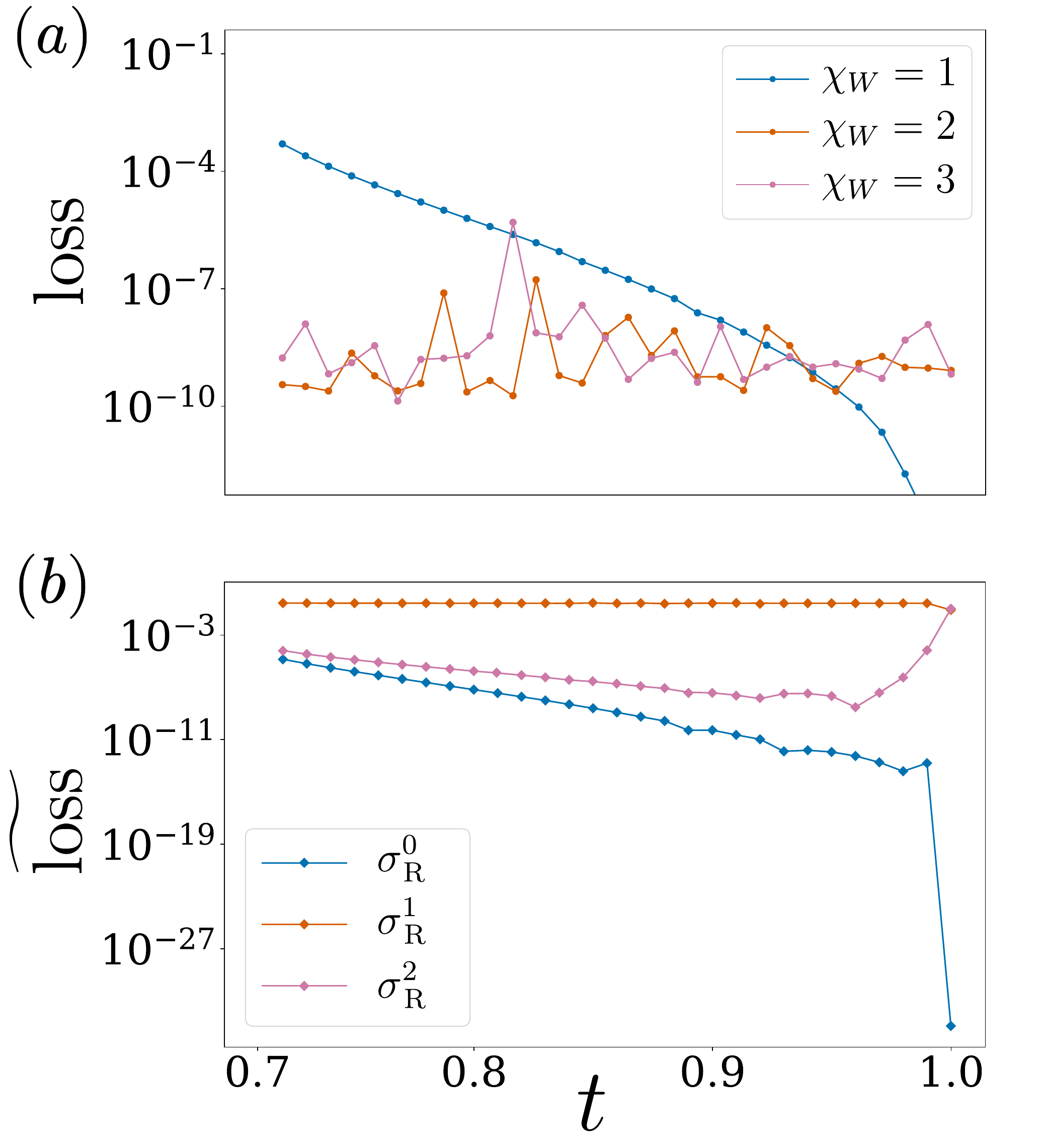}
    \caption{Results from gradient-descent optimization. (a) Loss function Eq.~\eqref{eq:loss_function_W} for each optimized solution $W_g$ solving the pulling-through relation across the interpolation (CZX phase only). We see that a bond-dimension-$1$ solution is only better very close to $t=1$. Away from that fixed point, solutions of higher bond dimension have smaller loss. (b) Test of the pulling-through relation $\ket{\sigma^g_\mathrm{R}}=\mathcal{W}_g\ket{\sigma_\mathrm{R}}$, where we use $W_g$ of the CZX fixed point, for a cylinder of width $L=3$. We use the loss function Eq.~\eqref{eq:loss_leading_eigvecs_symmetry_check}. The leading eigenvector ($i=0$) satisfies this relation better than subleading ones ($i=1,2$). These eigenvectors were obtained via exact diagonalization.}
    \label{fig:variational_loss_W_and_lead_eigvecs}
\end{figure}

In the 1D case, virtual symmetry representations $V_g$ are obtained as leading eigenvectors of the mixed TM $T_g$ [see Sec.~\ref{sec:1D_section}]. In 2D, while one may derive constraints on $W_g$ from the pulling-through relation~\cite{PhysRevB.94.205150, 10.1063/1.5007017}, we are unaware of generic equations that allow for the solution of $W_g$, e.g., as a simple linear algebra problem.  A direct numerical approach is to variationally solve for them. Consider minimizing the difference of both sides of the pulling-through relation [Fig.~\ref{fig:pulling_through_relation}(b)], i.e.
\begin{equation}
    \delta := AW_uW_l - U_g AW_rW_d,
\end{equation}
which is a rank-$5$ tensor of the same shape as the PEPS tensor $A$. Here $W_g$ is a genuine MPO of bond dimension $\chi_W$. Then, consider the loss function
\begin{equation}\label{eq:loss_function_W}
    \mathrm{loss} = |\delta|^2 / |A|^2,
\end{equation}
where $|X|$ denotes the Frobenius norm of a tensor $X$. We run gradient descent (specifically, the limited-memory Broyden–Fletcher–Goldfarb–Shanno algorithm~\cite{Liu1989LBFGS}) to seek a solution $W_g$ minimizing this loss, for the variationally optimized iPEPS tensor interpolation described in Sec.~\ref{sec:example}. Results are shown in Fig.~\ref{fig:variational_loss_W_and_lead_eigvecs}(a). We find that $W_g$ can indeed be solved for ``cleanly'' (with loss at machine precision) at renormalization group fixed points ($t=0$ and $t=1$), and for small deviations away from them. This is expected, as we analytically know $W_g$ for these fixed points [see Sec.~\ref{sec:example}]. As we deviate from fixed points, a bond-dimension-$1$ solution becomes increasingly inaccurate. Enlarging the internal bond dimension reduces the error. However, the degeneracy of $\tau_g$ (and hence the scaling of $\langle U^{x>0}_g \rangle$) is sensitive to small instabilities in the numerical solution for $W_g$. Similar instabilities were reported in~\cite{PhysRevB.98.125112, PhysRevB.82.165119} for PEPS representing topologically ordered ground states. Previous works resorted to imposing the virtual symmetries on the PEPS tensor~\cite{Sahinoglu2021, PhysRevB.98.165115}, which presupposes knowledge of these virtual symmetries. It also assumes a representation of arbitrary ground states in the phase diagram by a specific subclass of PEPS, i.e. MPO-injective PEPS, the only family for which the pulling-through relation is true in the first place. By contrast, the method outlined here only requires knowledge of the ground state PEPS tensor $A$ and the candidate physical symmetry representation $U_g$. While its interpretation is much aided by assuming the existence of such virtual symmetries $W_g$, it would be interesting to investigate the extent to which this method does not rely on the MPO-injective form.
 
While the loss is small away from the two renormalization-group fixed points, it being nonzero requires interpretation. This contrasts with the 1D case, where $V_g$ is obtained exactly as the leading eigenvector of the mixed TM $T_g$. In addition, with efficiency in mind, even if we accept this finite loss for the variational solution of $W_g$, it still represents by far the most time-consuming part of the detection algorithm. Realizing that $\tau_g$ (and hence $\langle U^{x>0}_g \rangle$) does not require explicit knowledge of $W_g$ drastically improves efficiency. 

On the other hand, let us test whether $\ket{\sigma^g_\mathrm{R}}=\mathcal{W}_g\ket{\sigma_\mathrm{R}}$ is true, with $\mathcal{W}_g$ obtained from the fixed-point CZX tensors. We define the second loss
\begin{equation}\label{eq:loss_leading_eigvecs_symmetry_check}
    \widetilde{\text{loss}} = |\ket{\sigma^g_\mathrm{R}} - \mathcal{W}_g\ket{\sigma_\mathrm{R}}|^2 / |A|^2.
\end{equation}
We plot this for various eigenvectors $\sigma^i_{\mathrm{R}}$, where eigenvalues are sorted in decreasing order (e.g. $i=0$ is the leading eigenvector), in Fig.~\ref{fig:variational_loss_W_and_lead_eigvecs}(b). 
The leading eigenvector of the PEPS TM $\mathcal{T}$ does retain the fixed point symmetry with a loss of order $10^{-11}$. This can be seen as a renormalization idea: deforming the state away from the CZX fixed point, the PEPS tensor itself does not retain the fixed-point symmetry, but a semi-infinite contraction of these does, as reflected in the leading eigenvectors. Importantly, subleading eigenvectors do not retain it, showing that the TM $\mathcal{T}$ itself does not have the fixed point symmetry.

\section{Transfer matrices at the CZX fixed point}\label{appendix:CZX_fixed_point}
\begin{figure}
    \centering
    \includegraphics[width=0.85\linewidth]{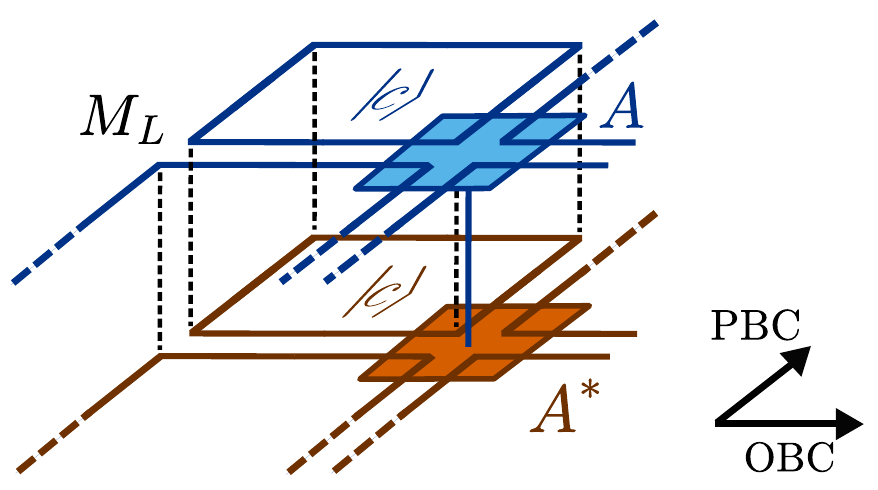}
    \caption{The leading eigenvectors of the transfer matrix for the CZX model. We use the explicit form of the PEPS for the CZX model in Fig.~\ref{fig:CZX_model_PEPS}, where pairs of virtual legs sharing a vertex form a MPS tensor for a cat state. On the right-hand side of the figure, the pair $A, A^*$ contracted along their physical leg form one rung of the transfer matrix along a cylinder $\mathcal{T}$ [see Fig.~\ref{fig:transfer_matrices_2D}]. When contracting those legs on their left with a similar tensor product of cat state MPSs, we obtain full overlaps of $4$-particle cat states on the bra and ket layers $\bra{c} c \rangle = 1$, which form full parallelepipeds. As a result we do have $\bra{\sigma_\mathrm{L}}\mathcal{T} = \bra{\sigma_\mathrm{L}}$ with the MPO $M_L$ for $\sigma_\mathrm{L}$ as shown. $M_L$ has bond dimension $\chi=4$ as its virtual leg is made of one ket and one bra leg of the MPS. Similar reasoning gives $\ket{\sigma_\mathrm{R}}$ when acting with a similar construction on the right-hand side of $\mathcal{T}$.}
    \label{fig:CZX_diagram_leading_eigvecs}
\end{figure}

\begin{figure}
    \centering
    \includegraphics[width=\linewidth]{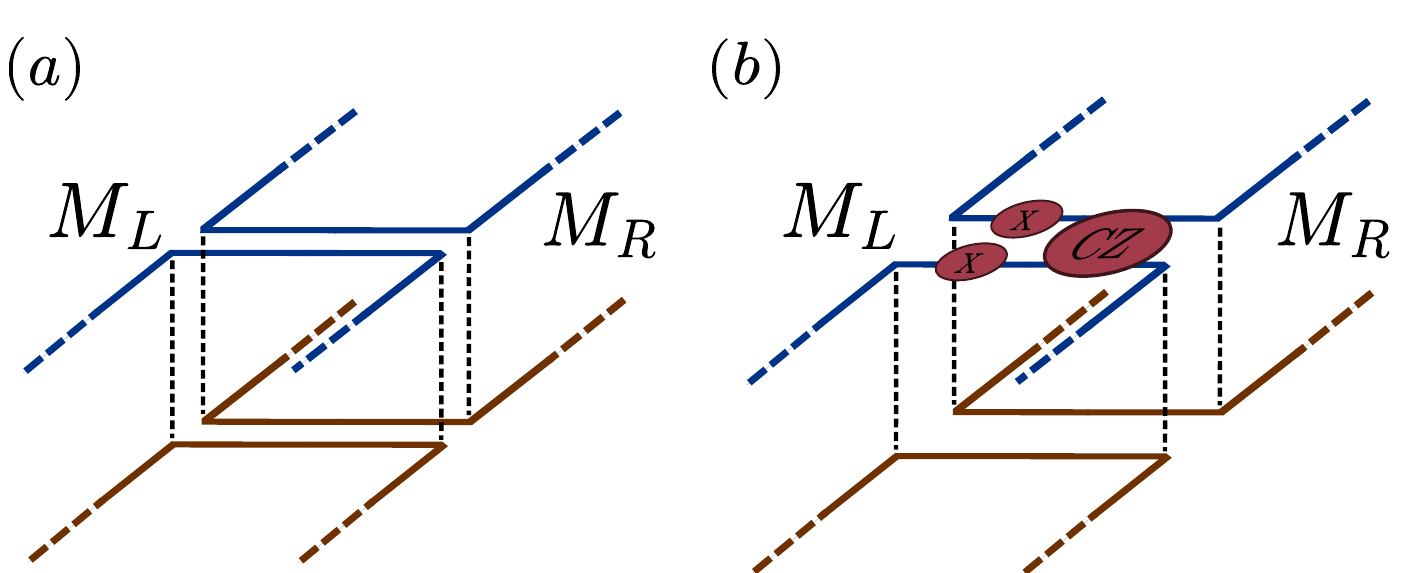}
    \caption{(a) $\tau$ for the CZX fixed point, where each pair of virtual legs sharing a vertex represents a cat-state MPS. $\tau$ has a single non-zero eigenvalue of $1$. (b) $\tau_g$ for the CZX fixed point, which is simply $\tau$ with an extra $W_g$ virtual matrix inserted.}
    \label{fig:CZX_diagram_tau_g}
\end{figure}

Using the PEPS construction for the CZX model [see Fig.~\ref{fig:CZX_model_PEPS}(a)], one can analytically determine the form of the leading eigenvectors of the PEPS TM. This is shown in Fig.~\ref{fig:CZX_diagram_leading_eigvecs}. 
The MPO $M_L$ for $\sigma_\mathrm{L}$ is simply a tensor product of two cat-state MPSs $A^i_{\alpha \beta} = \delta_{i=\alpha=\beta}$ on the ket layer, contracted along their physical legs with a tensor product of two other cat-state MPSs on the bra layer. In this fixed point case, $M_R$ is the same up to arbitrary orientation of the free virtual legs. One can then compute $\tau = \bra{\sigma_\mathrm{L}}\sigma_\mathrm{R}\rangle$ to see that $\mathrm{Tr}(\tau^L)$ is simply a multiplication of $L$ parallelepipeds of overlaps $\bra{c}c\rangle=1$, as expected. Secondly, at the CZX fixed point, the leading right eigenvector $\ket{\sigma^g_\mathrm{R}}$ of the mixed TM is simply $\mathcal{W}_g \ket{\sigma_\mathrm{R}}$ with local $W_g$ shown in Fig.~\ref{fig:CZX_model_PEPS}(b). We therefore know $\tau_g$ analytically, as shown in Fig.~\ref{fig:CZX_diagram_tau_g}.
We are interested in computing $\langle U^{x>0}_g \rangle = \mathrm{Tr}(\tau_g^L)$ for a given cylinder width $L$. From the figure, we see that the pair of $X$ Pauli gates on the ket layer yields orthogonal states to the bra layer. The $\CZ$ gates only affect these states by a sign, so contracting multiple $\tau_g$ instances gives $\langle U^{x>0}_g \rangle = 0$. This matches Ref.~\onlinecite{PhysRevB.90.235113}, where the fixed point of the Levin--Gu model also yields trivial diagnostics. This also matches the numerics, where the eigenvalues of $\tau_g$ tend to 0 as we approach the CZX fixed point (i.e. $t \rightarrow 1$). The same reasoning applies for the Levin--Gu version of the $\mathbb{Z}_2$ SPT~\cite{PhysRevX.5.041034, PhysRevB.95.125107}; the derivation above is analogous with hexagonal legs instead of square ones, giving $\langle U^{x>0}_g \rangle = 0$. At the CZX fixed point, the leading eigenvectors of Fig.~\ref{fig:CZX_diagram_leading_eigvecs} give the gauge tensor of Eq.~\eqref{eq:Vg_Vg*} explicitly. Each boundary MPS bond is a pair $(\kappa,\beta)$ of the ket- and bra-layer cat-state MPS bonds. Exchanging bra and ket in $M^g_\mathrm{R}$ exchanges the two bond factors; the mismatch of the $\CZ$ phase gives a $Z$ on one of them, so that
\begin{equation}
   V_g = (Z \otimes \id)\cdot \mathrm{SWAP}, \qquad V_g V^*_g = Z\otimes Z.
\end{equation}
Since $U_\CZX$ flips the spins, $M^g_\mathrm{R}$ only uses bond states with $\beta = 1-\kappa$, i.e. $\ket{01}$ and $\ket{10}$. This $\chi=4$ tensor is not injective, so $V_g$ is fixed only on this subspace, where $V_gV^*_g = -\id$, as expected and numerically confirmed in Fig.~\ref{fig:final_phase_diagram}(c). Restricted to it, $V_g$ reduces to $XZ$ on the injective $\chi=2$ boundary MPS.

\section{Technical proofs}
We here prove the remaining statements required by the main text.
\subsection{Proof that $V_{\mathrm{L}}^{\mathrm{H}} V_{\mathrm{L}}^{\mathrm{H}*}=
V_{\mathrm{R}}^{\mathrm{H}} V_{\mathrm{R}}^{\mathrm{H}*}=1
$}\label{app:VL_VL*}
Here we show that the gauge matrices $V_{\mathrm{L}/\mathrm{R}}^{\mathrm{H}}$ of the Hermitian conjugation operation satisfy Eq.~\eqref{eq:Hermitian_gauge}, which is rewritten here:
\begin{equation}
V^{\mathrm{H}}_{\mathrm{L}} V^{\mathrm{H}*}_{\mathrm{L}} = \mathbb 1, \quad V_{\mathrm{R}}^{\mathrm{H}} V_{\mathrm{R}}^{\mathrm{H}*} = \mathbb 1.
\end{equation}
Using Eq.~\eqref{eq:env_Hermitian_eq} twice, we obtain
\begin{equation}
\begin{aligned}
    M_{\mathrm{L}}^{ji} &= 
     V_{\mathrm{L}}^{\mathrm{H}} V_{\mathrm{L}}^{\mathrm{H}*} M_{\mathrm{L}}^{ji}
     (V_{\mathrm{L}}^{\mathrm{H}} V_{\mathrm{L}}^{\mathrm{H}*})^{-1}, \\
    M_{\mathrm{R}}^{ij} &= 
     V_{\mathrm{R}}^{\mathrm{H}} V_{\mathrm{R}}^{\mathrm{H}*} M_{\mathrm{R}}^{ij}
     (V_{\mathrm{R}}^{\mathrm{H}} V_{\mathrm{R}}^{\mathrm{H}*})^{-1}
\end{aligned}
\end{equation}
for all $i,j$.
Since $M_{\mathrm{L/R}}$ are injective, this implies
\begin{equation}
    V_{\mathrm{L}}^{\mathrm{H}} V_{\mathrm{L}}^{\mathrm{H}*} = \lambda_{\mathrm{L}} \mathbb{1}, \quad
    V_{\mathrm{R}}^{\mathrm{H}} V_{\mathrm{R}}^{\mathrm{H}*} = \lambda_{\mathrm{R}} \mathbb{1},
\end{equation}
with $\lambda_{\mathrm{L/R}} \neq 0$ since $V_{\mathrm{L/R}}^{\mathrm{H}}$ are invertible.
Suppressing the labels, $VV^* = \lambda\mathbb 1$ gives $V^* = \lambda V^{-1}$, hence
$V^*V = \lambda\mathbb 1$. Complex conjugation yields $VV^* = \lambda^*\mathbb 1$, and
comparing with $VV^* = \lambda\mathbb 1$ we find $\lambda \in \mathbb R$.
Rescaling $V_{\mathrm{L/R}}^{\mathrm{H}} \to |\lambda_{\mathrm{L/R}}|^{-1/2} V_{\mathrm{L/R}}^{\mathrm{H}}$,
which leaves Eq.~\eqref{eq:env_Hermitian_eq} invariant, we can set $\lambda_{\mathrm{L/R}} = \pm 1$.

To fix the signs, we define $t_{\mathrm{L}} = \sum_{i} M_{\mathrm{L}}^{ii}$ and $t_{\mathrm{R}} = \sum_i M_{\mathrm{R}}^{ii}$. Equation~\eqref{eq:env_Hermitian_eq} implies
\begin{equation}\label{eq:t_anti_unitary}
    t_{\mathrm{L}} = V_{\mathrm{L}}^{\mathrm{H}} t^*_{\mathrm{L}}  (V_{\mathrm{L}}^{\mathrm{H}})^{-1}, \ 
    t_{\mathrm{R}} = V_{\mathrm{R}}^{\mathrm{H}} t^*_{\mathrm{R}}  (V_{\mathrm{R}}^{\mathrm{H}})^{-1}.
\end{equation}
We assume that $t_{\mathrm{L/R}}$ has a unique leading eigenvalue, 
\begin{equation}
    t_{\mathrm{L}} |\mu_{\mathrm{L}} \rangle = \mu_{\mathrm{L}} | \mu_{\mathrm{L}} \rangle, \quad
    t_{\mathrm{R}} |\mu_{\mathrm{R}} \rangle = \mu_{\mathrm{R}} | \mu_{\mathrm{R}} \rangle,
\end{equation}
with $\mu_{\mathrm{L/R}}$ the leading eigenvalue. Then $\mu_{\mathrm{L}}, \mu_\mathrm{R} \in \mathbb{R}$ by Eq.~\eqref{eq:t_anti_unitary}, and 
\begin{equation}
     V_{\mathrm{L}}^{\mathrm{H}} |\mu_{\mathrm{L}}^*\rangle = 
    N_{\mathrm{L}}|\mu_{\mathrm{L}}\rangle , \quad
     V_{\mathrm{R}}^{\mathrm{H}} |\mu_{\mathrm{R}}^*\rangle = 
    N_{\mathrm{R}}|\mu_{\mathrm{R}}\rangle,
\end{equation}
where $N_{\mathrm{L}}$ and $N_{\mathrm{R}}$ are the proportionality constants.
Applying $V^{\mathrm{H}}_{\mathrm{L/R}}\mathcal{K}$ ($\mathcal{K}$ is complex conjugation) again, we obtain 
\begin{equation}
     V_{\mathrm{L}}^{\mathrm{H}}V_{\mathrm{L}}^{\mathrm{H}*} |\mu_{\mathrm{L}}\rangle = 
    |N_{\mathrm{L}}|^2 |\mu_{\mathrm{L}}\rangle , \,
     V_{\mathrm{R}}^{\mathrm{H}}V_{\mathrm{R}}^{\mathrm{H}*} |\mu_{\mathrm{R}}\rangle = 
    |N_{\mathrm{R}}|^2 |\mu_{\mathrm{R}}\rangle .
\end{equation}
Therefore, both signs must be positive $\lambda_{\mathrm{L}} = \lambda_\mathrm{R} = 1$, and Eq.~\eqref{eq:Hermitian_gauge} must hold.

\subsection{Injective MPS representation of $\sigma_{\mathrm{R}}\mathcal W_g$}
\label{app:injective_sigma_R_Wg}

Given that $\sigma_{\mathrm R}$ has an injective MPS representation
$M_{\mathrm R}$, we prove that
$\sigma_{\mathrm R}\mathcal W_g$ admits an injective representation after finite blocking. All MPO families below are generated by fixed local tensors under periodic boundary conditions.

Denote
$C_N=\sigma_{\mathrm R,N}\mathcal W_{g,N}$, where $N$ is the length of the MPO.
After a finite blocking, its canonical decomposition gives
\begin{equation}
    C_N=\sum_a\sum_{j=1}^{A_a}\mu_{a,j}^{N}\mathcal W_N(B_a),
\end{equation}
where $a$ labels the equivalence classes of injective tensors,
$B_a$ is a representative of each class, and $j$ labels
equivalent copies. We omit zero coefficients, so
$\mu_{a,j}\neq0$.
Using the supported group relations, we obtain
\begin{equation}
\begin{aligned}
    C_N\mathcal W_{g^{-1},N}
    &=\sigma_{\mathrm R,N} \mathcal W_{g,N}\mathcal W_{g^{-1},N}\\
    &=\sigma_{\mathrm R,N}P_N=\sigma_{\mathrm R,N},
\end{aligned}
\end{equation}
where we have used $\sigma_{\mathrm{R}, N}P_N =  \sigma_{\mathrm{R}, N}$.

Consequently,
\begin{equation}
    \sum_a\sum_{j=1}^{A_a}
    \mu_{a,j}^{N} \mathcal W_N(B_a)\mathcal W_{g^{-1},N} = \sigma_{\mathrm R,N}.
\end{equation}

To apply block independence, we first canonically decompose
each product
$\mathcal W_N(B_a)\mathcal W_{g^{-1},N}$.
After further blocking, write
\begin{equation}
    \mathcal W_N(B_a)\mathcal W_{g^{-1},N}
    =
    \sum_c\sum_{k\in K_{ac}}
    \eta_{a,c,k}^{N}\mathcal W_N(Q_c),
\end{equation}
where $\eta_{a,c,k}\neq0$ and the injective tensors $Q_c$
are pairwise inequivalent up to an auxiliary gauge and a
nonzero per-site scalar. Include $Q_0=M_{\mathrm R}$ in
this collection.

Substituting these decompositions gives
\begin{equation}
    \sum_c
    \left[
        \sum_a\sum_{j=1}^{A_a}\sum_{k\in K_{ac}}
        (\mu_{a,j}\eta_{a,c,k})^{N}
    \right]
    \mathcal W_N(Q_c)=\mathcal W_N(Q_0).
\end{equation}
Linear independence of inequivalent normal MPS for sufficiently
large $N$ therefore implies
\begin{equation}
    \sum_a\sum_{j=1}^{A_a}\sum_{k\in K_{ac}}
    (\mu_{a,j}\eta_{a,c,k})^{N}
    = \delta_{c0}.
\end{equation}
By uniqueness of finite power sums~\cite{10.1063/1.5007017},
there are no nonzero terms for $c\neq0$, while precisely
one term remains for $c=0$. Relabeling its indices as
$(a,j,k)=(1,1,1)$, we have
$\mu_{1,1}\eta_{1,0,1}=1$.

Since every $\mu_{a,j}$ is nonzero, the surviving
$\eta_{1,0,1}$ would contribute one nonzero term for
each $j=1,\ldots,A_1$. Uniqueness of the surviving term
therefore also implies $A_1=1$. Thus
\begin{equation}\label{eq:block_indep}
\begin{aligned}
    A_1&=1,\\
 \mu_{1,1}\eta_{1,0,1}&=1,\\
    \mu_{1,1}^{N}\mathcal W_N(B_1)
        \mathcal W_{g^{-1},N}
        &=\sigma_{\mathrm R,N},\\
    \mathcal W_N(B_a)\mathcal W_{g^{-1},N}
        &=0,\qquad a>1.
\end{aligned}
\end{equation}
These identities hold for every length in the common
blocked unit cell.

Multiplying from the right by $\mathcal W_{g,N}$ gives
\begin{equation}
\begin{aligned}
    \mu_{1,1}^{N}\mathcal W_N(B_1)P_N&=C_N,\\
    \mathcal W_N(B_a)P_N&=0,\qquad a>1.
\end{aligned}
\end{equation}
Moreover, $C_NP_N=C_N$. Since $P_N=P_N^\dagger$, for
$a>1$ we have
\begin{equation}
\begin{aligned}
    \mathrm{Tr}\!\left(C_N^\dagger\mathcal W_N(B_a)\right)
    &=\mathrm{Tr}\!\left((C_NP_N)^\dagger\mathcal W_N(B_a)\right)\\
    &=\mathrm{Tr}\!\left(C_N^\dagger\mathcal W_N(B_a)P_N\right)=0.
\end{aligned}
\end{equation}

Define the mixed transfer matrix
\begin{equation}
    E_{b,a}
    :=
    \sum_{i,k}B_b^{*ik}\otimes B_a^{ik}.
\end{equation}
Keeping the scalar factors from the canonical decomposition,
the same overlap can be written as
\begin{equation}
\begin{aligned}
    0
    &=
    \mathrm{Tr}\!\left(C_N^\dagger\mathcal W_N(B_a)\right)\\
    &=\sum_b\sum_{j=1}^{A_b} (\mu_{b,j}^{*})^{N}\,
    \mathrm{tr}(E_{b,a}^{N})\\
    &= \mathrm{Tr}\!\left[
        \left( \bigoplus_b\bigoplus_{j=1}^{A_b} \mu_{b,j}^{*}E_{b,a}\right)^{N} \right].
\end{aligned}
\end{equation}
Since these power traces vanish for every $N$, the
direct-sum matrix has only zero eigenvalues.
However, it contains the blocks
$\mu_{a,j}^{*}E_{a,a}$ for $a>1$.
Injectivity of $B_a$ implies that $E_{a,a}$ has
a non-zero leading eigenvalue.
Thus these blocks have nonzero eigenvalues, a contradiction.

Therefore no blocks with $a>1$ remain. Absorbing $\mu_{1,1}$ into $B_1$, we obtain an injective representation
\begin{equation}
    \sigma_{\mathrm R,N}\mathcal W_{g,N}
    = C_N = \mathcal W_N(B_1).
\end{equation}

\subsection{Proof that $V_g V_g^*= \omega(g,g,g) \mathbb{1}$}\label{app:VgR_VgR*}

To establish $V_gV_g^*=\omega(g,g,g)\mathbb{1}$,
we first determine the Hermiticity sign of the symmetry MPO.
For the nontrivial element $g\in\mathbb{Z}_2$, supported
unitarity implies
\begin{equation}
    \mathcal W_g^\dagger=\mathcal W_{g^{-1}}=\mathcal W_g,
    \qquad
    \mathcal W_g^2=P,
\end{equation}
where $\mathcal W_e=P$ is the orthogonal projector onto the
relevant virtual support. Thus $\mathcal W_g$ is Hermitian,
although it need not be unitary on the full virtual boundary space.
Given that the MPO tensor $W_g$ is injective after finite
blocking, the fundamental theorem gives the local
Hermiticity gauge
\begin{equation}\label{eq:MPO_inverse}
    \includegraphics[height=0.1\textheight]
        {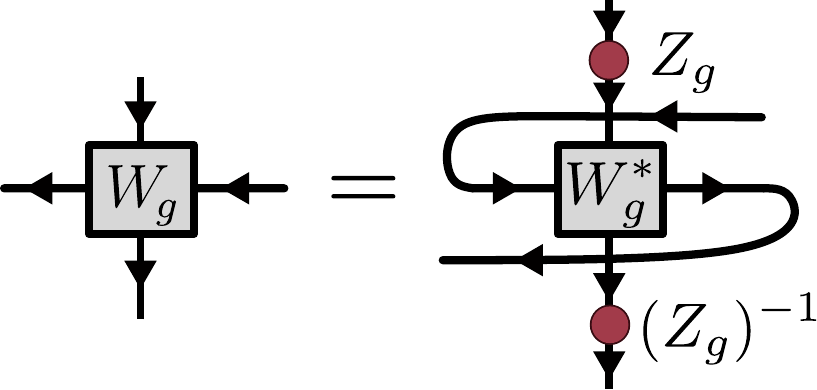}.
\end{equation}
Applying this relation twice and using injectivity shows
that $Z_gZ_g^*$ is a nonzero real scalar multiple of the
identity. After rescaling $Z_g$, we may therefore write
\begin{equation}
    Z_gZ_g^*=\nu_g\mathbb{1},
    \qquad \nu_g\in\{+1,-1\}.
\end{equation}

We invoke the Frobenius--Schur relation stated in
Ref.~\onlinecite{bultinck_2017}, Appendix~A, Eqs.~(136)--(138),
for Hermitian projector-MPO data with a simple unit,
the unital spectral property, zipper identities, and
coherent unit fusion. 
In our convention the splitting $F$-symbol is
$\omega(g,g,g)^{-1}$, and the identity reads
\begin{equation}
    [\mathsf F^{ggg}_g]_{ee} =\omega(g,g,g)^{-1} =\nu_g.
\end{equation}
Therefore,
\begin{equation}\label{eq:Zg_Zg*}
    Z_gZ_g^*=\omega(g,g,g)^{-1}\mathbb{1}.
\end{equation}

To relate the MPO sign Eq.~\eqref{eq:Zg_Zg*} to the boundary gauge, define the transfer matrix $m$ whose periodic contractions generate $\mathrm{Tr}(\sigma_{\mathrm R}\mathcal W_g^2)$
\begin{equation}
    m = \sum_{ij}  W_g^{ji} \otimes (M_{\mathrm{R}}^g)^{ij},
\end{equation}
where $i$ and $j$ are the physical indices of the MPOs.
$m$ has an anti-linear symmetry
\begin{equation}\label{eq:Wg_MgR_proof}
    \includegraphics[height=0.1\textheight]{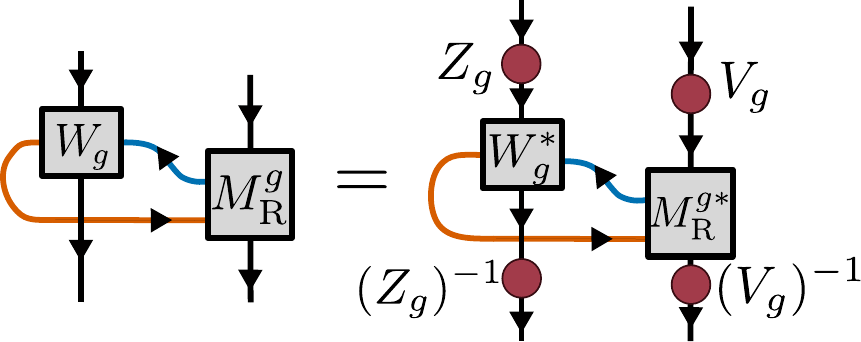}.
\end{equation} 
By the same logic used in Appendix~\ref{app:VL_VL*}, 
\begin{equation}
    (Z_g \otimes V_g) (Z_g^* \otimes V_g^*) = \epsilon_g \mathbb{1}, 
\end{equation}
where $\epsilon_g$ can be $\pm 1$.
To determine $\epsilon_g$, we note that $\mathrm{Tr}(
\sigma_{\mathrm{R}} \mathcal{W}_g^2 ) = \mathrm{Tr}(\sigma_{\mathrm{R}})$, meaning that the transfer matrix $m$ has the same non-zero eigenvalue spectrum as $t_{\mathrm{R}}$. Therefore, $m$ also has the same unique leading eigenvalue as $t_{\mathrm{R}}$. Using the same argument as in Appendix~\ref{app:VL_VL*}, we get 
\begin{equation}
    (Z_g \otimes V_g) (Z_g^* \otimes V_g^*) = \mathbb{1}.
\end{equation}
Combined with Eq.~\eqref{eq:Zg_Zg*}, we obtain that
\begin{equation}
    V_g V_g^* = \omega(g,g,g)\mathbb{1}.
\end{equation}

\section{Determining the gauge matrix $V_g$}
\label{app:Get_Vg}
We describe the procedure used to determine $V_g$ in Eq.~\eqref{eq:sigma_R_W_g_mpo}. 
In practice, the tensor returned by CTMRG may carry an unfixed overall phase. We therefore allow a phase $\eta\in U(1)$ in its local Hermiticity relation
\begin{equation}\label{eq:MgR_U1}
    (M_{\mathrm R}^{g})^{ij}
    = \eta V_g (M_{\mathrm R}^{g})^{*ji}V_g^{-1}.
\end{equation}
Consider the transfer matrices
\begin{equation}
\begin{aligned}
    E &= \sum_{i,j} (M_{\mathrm{R}}^g)^{*ji} \otimes (M_{\mathrm{R}}^g)^{ji}, \\
    \tilde{E} &= \sum_{i,j} (M_{\mathrm{R}}^g)^{ij}
    \otimes (M_{\mathrm{R}}^g)^{ji}.
\end{aligned}
\end{equation}
By the gauge relation Eq.~\eqref{eq:MgR_U1}, 
\begin{equation}
    \tilde{E} = \eta(V_g \otimes \mathbb1) E (\ V_g^{-1} \otimes \mathbb 1).
\end{equation}
Denote the right leading eigenvectors of $E$ and $\tilde{E}$ as $|\rho\rangle$ and $|\tilde{\rho}\rangle$, respectively:
\begin{equation}
    E |\rho \rangle= \mu |\rho\rangle, \quad \tilde{E} |\tilde{\rho}\rangle = \tilde{\mu} |\tilde \rho \rangle.
\end{equation}
We then have
\begin{equation}
    \tilde{\mu} = \eta {\mu}, \quad |\tilde{\rho}\rangle = c ( V_g \otimes \mathbb1)|\rho\rangle,
\end{equation}
where $c$ is a scalar constant.
We define the matricization mapping
\begin{equation}
|X\rangle = \sum_{ab} X_{ab} |a\rangle \otimes|b\rangle \to X_{ab}
\end{equation}
and denote matricizations of $|\tilde{\rho}\rangle$ and $|\rho\rangle$ as $\tilde \rho$ and $\rho$, respectively.
Because $M^g_{\mathrm{R}}$ is injective after finite blocking,
the matricized leading eigenvector $\rho$ is invertible. Therefore, we obtain that
\begin{equation}
   \tilde {\rho}  = c V_g \rho \implies V_g = c^{-1}\tilde \rho \rho^{-1}.
\end{equation}
The undetermined scalar $c$ corresponds to the multiplicative scalar gauge freedom of $V_g$ in Eq.~\eqref{eq:MgR_U1}, and we fix it by requiring that
\begin{equation}
    V_g V_g^* = \omega(g,g,g) \mathbb{1},
\end{equation}
where $\omega(g,g,g) \in \{+1, -1\}$ is the 3-cocycle.

\end{document}